\documentclass[fleqn,10pt]{wlscirep}
\usepackage[utf8]{inputenc}
\usepackage[T1]{fontenc}
\usepackage{siunitx}
\usepackage{graphicx}
\usepackage{float}
\usepackage{caption}
\usepackage{subcaption}
\usepackage{comment}
\usepackage{mathtools}
\usepackage{balance}
\usepackage{multirow}
\usepackage{lineno}

\title{Design of Energy-Efficient Cross-coupled Differential Photonic-SRAM (pSRAM) Bitcell for High-Speed On-Chip Photonic Memory and Compute Systems}
\author[1,*]{Md Abdullah-Al Kaiser}
\author[2]{Sugeet Sunder}
\author[2]{Clynn Mathew}
\author[3]{Michal Rakowski}
\author[2]{Ajey P. Jacob}
\author[1]{Akhilesh R. Jaiswal}
\affil[1]{University of Wisconsin--Madison, Madison, Wisconsin, United States}
\affil[2]{Information Sciences Institute, University of Southern California, Los Angeles, California, United States}
\affil[3]{GlobalFoundries, Malta, New York, United States}
\affil[*]{mkaiser8@wisc.edu}


\begin{abstract}
In this work, we propose a novel differential photonic static random access memory (pSRAM) bitcell design using fabrication-friendly photonic components. The proposed pSRAM overcomes the key limitations of traditional electrical SRAMs, which struggle with speed and power efficiency due to increasing bitline/wordline capacitance and interconnect resistance associated with long electrical wires as technology scales. By utilizing cross-coupled micro-ring resonators and differential photodiode structures, along with optical waveguides instead of traditional wordlines and bitlines, our pSRAM exhibits high-speed, and energy-efficient performance. The pSRAM bitcell demonstrates a read/write speed of 40 GHz, with a switching (static) energy consumption of approximately 0.6 pJ (0.03 pJ) per bit and a footprint of \si{330 \times 290} \si{\micro m^2} using the GlobalFoundries 45SPCLO process node. These bitcells can be arranged into a 2D memory array, enabling large-scale, on-chip photonic memory subsystems ideal for high-speed memory, data processing and computing applications.

\end{abstract}
\begin{document}

\flushbottom
\maketitle

\thispagestyle{empty}

\section{Introduction}

The rising demand for high-speed and energy-efficient computing has highlighted the limitations of conventional electronic memory technologies. As transistor technology scales, electronic memory faces increasing challenges, fundamentally limited by the increased bitline and wordline interconnect capacitance and the resistance of long electrical interconnects \cite{eSRAM_prb}. These issues considerably reduce speed, bandwidth and power efficiency of the memory system, leading to what is known as the ``memory wall bottleneck'' \cite{memory_wall_bottleneck}. Although advancements in optical communication have led to improvements in data transfer rates and power efficiency, electronic memory systems remain constrained by limited bandwidth and extended access times, ultimately limiting modern computing systems' processing capabilities \cite{optics_advantage1, optics_advantage2, optics_advantage3}. This problem is particularly pronounced with the rise of data-intensive applications such as artificial intelligence, machine learning, and large-scale simulations, all of which necessitate rapid and efficient memory solutions \cite{dl_ref, tpu_ref}. In this regard, photonic memory technologies emerge as a promising alternative \cite{optical_ram_survey}, offering high bandwidth, ultra-fast access speeds and minimal interference, which can potentially circumvent the limitations posed by the electrical systems. However, developing static, compact, ultra-fast, energy-efficient, and CMOS compatible photonic memory that meets these demands remains a substantial challenge. 
Overcoming these challenges is essential to pave the pathway for widespread adoption of photonic memory systems in communication and computing applications, providing significant performance, bandwidth and energy efficiency improvements.  

Various photonic memory implementations have been explored using methods like semiconductor optical amplifier-based Mach-Zehnder interferometers (SOA-MZIs) \cite{SOA-MZI1, SOA-MZI2}, semiconductor optical amplifier cross-gain modulation (SOA-XGM) switches \cite{SOA-XGM1, SOA-XGM2}, and SOA-based coupled ring lasers \cite{SOA-Ring_Laser}, which utilize wavelength selectivity for data storage. The SOA-based designs face significant challenges, particularly in terms of energy consumption and physical footprint. The energy required for SOA biasing (e.g., 250 mA \cite{SOA-MZI2}, 300 mA \cite{SOA-XGM2}) and optical state switching is considerable (e.g., 7.96 pJ\cite{SOA-MZI2}, 237.5 pJ \cite{SOA-XGM2}), and the footprint requirements often exceed a few mm\si{^2} (e.g., 12 mm\si{^2}\cite{SOA-MZI2}), thereby potentially limiting their practical application in compact, energy-efficient integrated systems. Micro-ring laser-based technology leverages the clockwise (CW) and counterclockwise (CCW) propagation of light to create bistable optical memory designs \cite{mr_laser1, mr_laser2, mr_laser3}. However, these laser-based memory technologies require a DC current bias (e.g., 30 mA \cite{mr_laser1}, 200 mA \cite{mr_laser2}) to fine-tune the resonant frequencies of the two lasers to be close to each other, resulting in high energy consumption due to the required bias current. Additionally, they rely on multiple quantum well (MQW) technology \cite{mr_laser2, mr_laser3}, necessitating significant modifications to existing standard CMOS foundry processes.

Micro-disk lasers \cite{disk_laser}, for instance, utilize clockwise and counterclockwise lasing modes as state variables, achieving low switching power but necessitating thermal tuning to maintain stability. Additionally, ultra-low power optical RAM based on photonic crystal nanocavities with ultra-compact buried heterostructures \cite{nanocavity1, nanocavity2, nanocavity3} has been developed, but these designs suffer from extended switch-off times (in the range of a few nanoseconds) due to slow carrier relaxation within the cavity. Furthermore, III-V-on-Si photonic crystal nanocavity laser-based optical memory \cite{iii_v_nanocavity1} supports high-speed operations but requires substantial modifications to existing silicon fabrication processes. Phase Change Memory (PCM)-based optical memories provide advantages like a compact footprint, multi-bit storage capability, and non-volatility; however, they also encounter challenges such as high write energy and limited speed, potentially restricting their deployment in high-speed environments \cite{pcm_ram1, pcm_ram2, pcm_ram3}. In summary, technologies such as SOA-MZI setups, ring and micro-disk lasers, photonic crystal nanocavities, and PCM-embedded waveguides face issues like high static/switching energy consumption, limited scalability, or the need for significant process alterations, limiting their integration potential for large-scale applications.

Hence, in this work, we introduce a novel differential photonic-SRAM (pSRAM) bitcell, designed using cross-coupled micro-ring resonators (MRRs) and photodiodes (PDs). The pSRAM retains data as long as the optical and electrical biases are maintained, exhibiting a true static behavior similar to traditional electrical SRAMs.  It supports differential write and read operations, storing both the data and its complement, thus exhibiting functional equivalence to its electrical counterparts. The pSRAM cells can be configured into a 2D array, forming a scalable on-chip memory architecture where any bitcell can be accessed through waveguide wordlines and bitlines. The exclusive use of matured photonic components (MRRs and PDs) makes the design suitable for large-scale manufacturing in existing silicon photonics foundries and also compatible with conventional electrical systems for monolithic integration. Our design has been validated using simulation based on state-of-the-art monolithic 45nm silicon photonics platform - the GlobalFoundries 45SPCLO technology, achieving ultra-fast read/write speeds of up to 40 GHz with a switching (static) energy of 0.6 pJ (0.03 pJ) per bit considering a wall-plug efficiency of 20\%, while occupying a compact area of \si{\sim}0.1 mm\si{^2} per bitcell.

\section{Methods}

This section covers various configurations of photonic cross-coupled latches, their operating principles, the process of writing data into the pSRAM bitcell, and the methods for reading data from it. In memory systems, a cross-coupled latch retains stored data as long as the bias is maintained, or until new data needs to overwrite it. The write operation allows new data to be stored in the latch, while the read operation retrieves this stored data for further processing. These fundamental hold, write, and read functions are critical for the functionality and stability of any memory system. Consequently, we discuss the robust and reliable operation principles of the proposed photonic pSRAM in detail here.

\subsection{Hold Operation of the photonic-SRAM (pSRAM) Bitcell}

\begin{figure}[!b]
\includegraphics[width=1\linewidth]{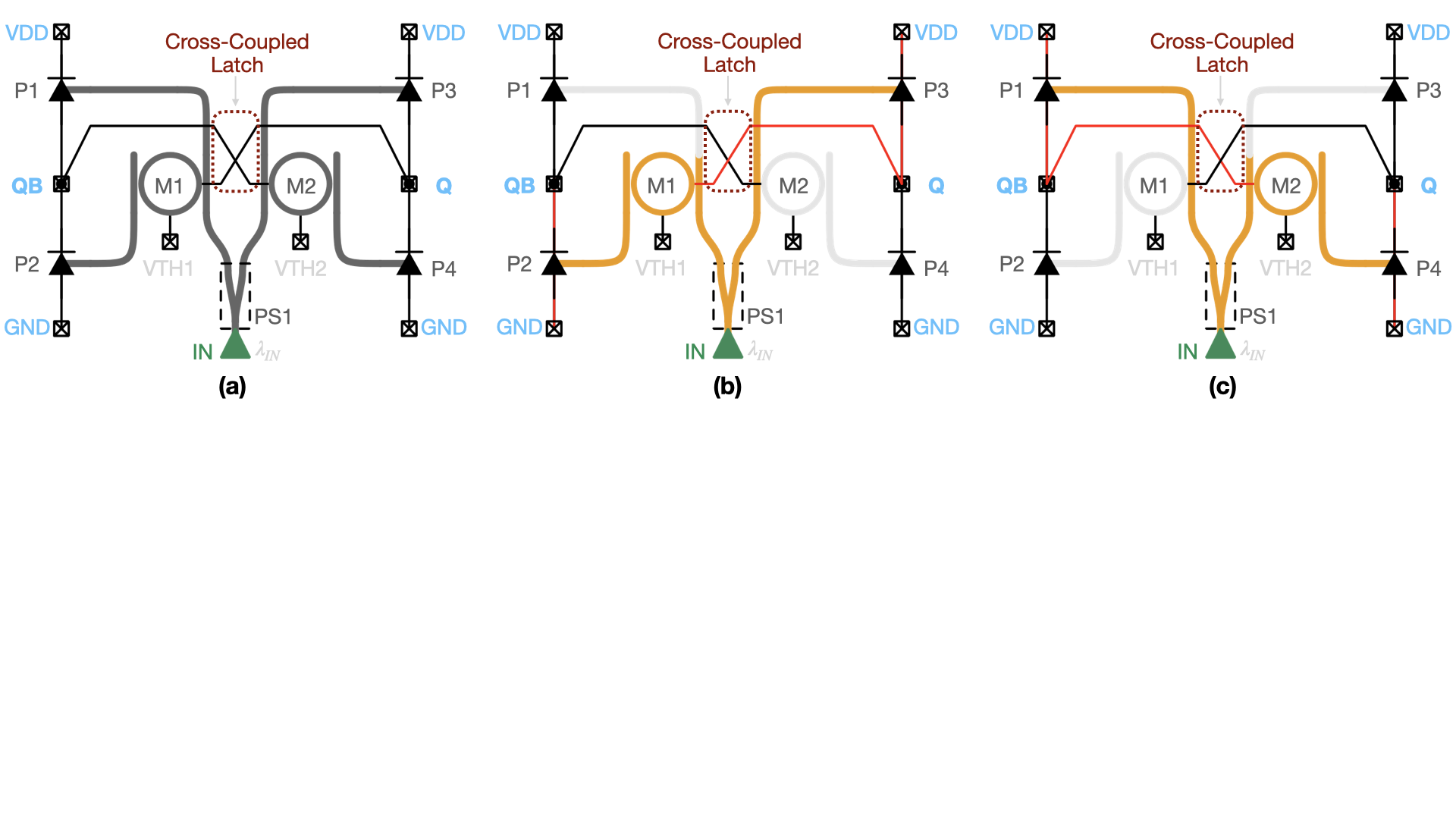}
\vspace{-55 mm}
\caption{(a) pSRAM bitcell hold structure featuring a cross-coupled micro-ring resonator configuration, demonstrating the hold operation of the pSRAM latch for (b) Q = 1, QB = 0, and (c) Q = 0, QB = 1, respectively. Thin lines indicate electrical wires, while thick lines depict photonic waveguides. Photonic waveguides shown in orange represent active light pathways, whereas light gray waveguides indicate inactive (dark) pathways, and the red electrical wires denote active connections in (b) and (c), respectively.}
\label{psram_latch_config1}
\end{figure}

\begin{figure}[!t]
\includegraphics[width=1\linewidth]{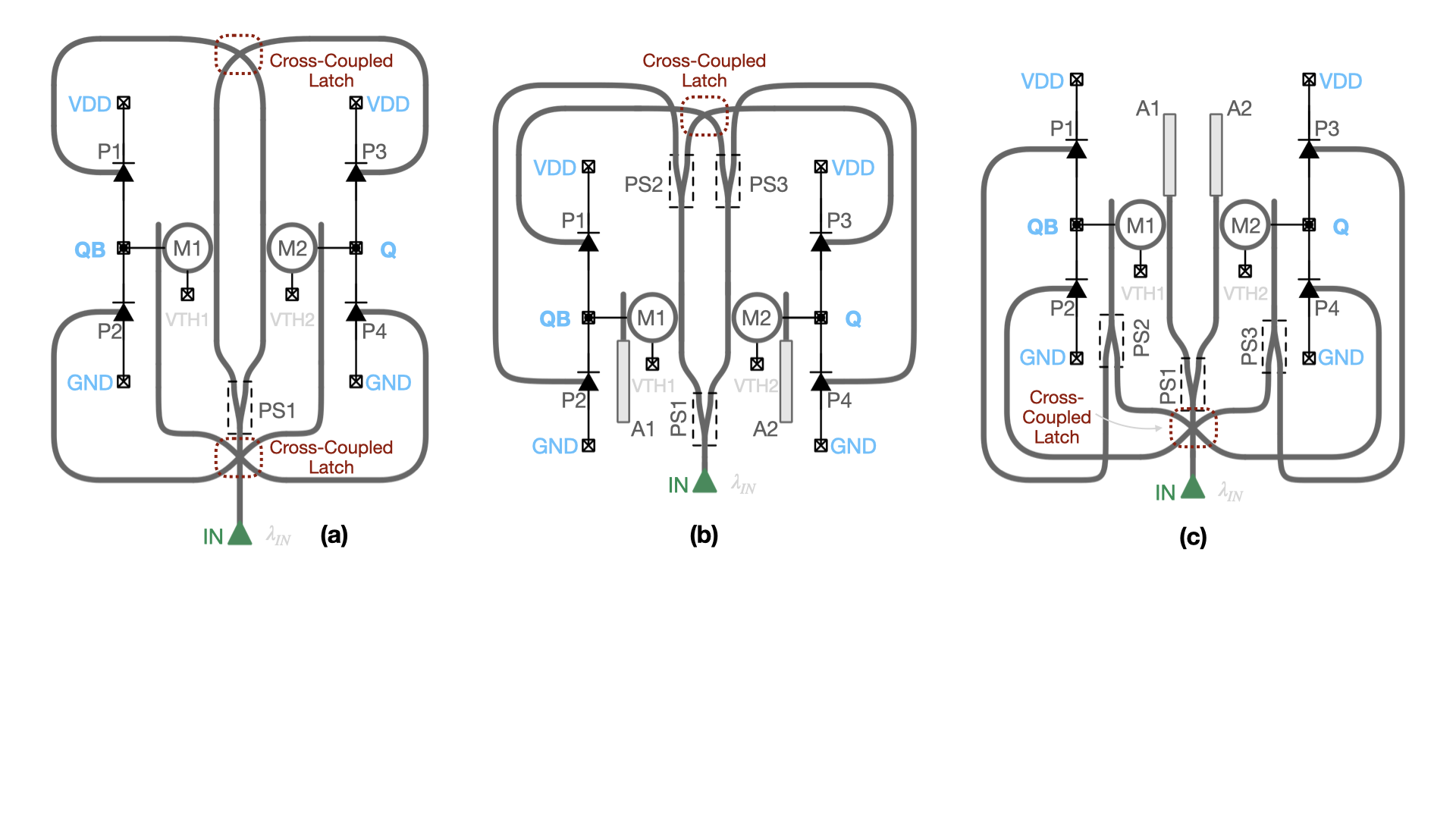}
\vspace{-35 mm}
\caption{pSRAM latch structure with cross-coupled (a) photodiode drive, (b) photodiode drive with through-port only, and (c) photodiode drive with drop-port only.}
\label{psram_latch_config2}
\end{figure}

Figure \ref{psram_latch_config1} and \ref{psram_latch_config2} show the various electro-optical latch configurations for the pSRAM bitcell. In the diagram, M1-M2 represent micro-ring resonators (MRRs), P1-P4 are photodiodes (PDs), PS1-PS3 are optical 50:50 power splitters, and A1-A2 are passive optical absorbers. An optical laser (\si{\lambda_{IN}}) is connected to the input power splitter (PS1) through the input port (\textbf{IN}), which delivers power to the input ports of two identical MRRs (M1 and M2). The wavelength \si{\lambda_{IN}} is chosen so that when a voltage of VDD is applied to MRRs M1 and M2, they resonate with the incoming light. The midpoints between photodiodes P1 and P2 (P3 and P4) are labeled as QB (Q), serving as the electrical storage nodes of the pSRAM. The operating principles of each configuration are discussed below.

\subsubsection{Cross-coupled Micro-ring Resonator Drive} \label{sec_latch_mrr_drive}
In the configuration illustrated in Figure \ref{psram_latch_config1}(a), the through and drop ports of M1 (M2) are connected to the waveguides of photodiodes P1 (P3) and P2 (P4), respectively. The node QB (Q) drives M2 (M1), forming a cross-coupled structure to store data. To maintain data at the storage nodes, the cross-coupled electro-optic structure must remain stable. For example, when Q = 1 (VDD) and QB = 0 (GND) are stored, these values are held as long as both optical and electrical biases are applied. Since Q = 1, M1 resonates with the input light, directing most of it to P2, generating higher current, which creates a low-resistance path to GND and holds QB at 0. In turn, QB keeps M2 out of resonance, allowing light to pass to P3, which maintains Q at VDD (as shown in Figure \ref{psram_latch_config1}(b)). Likewise, when QB = 1, M2 resonates with the incoming light, directing most of it to P4, keeping Q closer to GND. As a result, M1 is out of resonance, directing most light to P1 utilizing the through port to M1, maintaining QB at VDD (as shown in Figure \ref{psram_latch_config1}(c)). This cross-coupled MRRs and photodiodes structure ensures that the stored data remains latched as long as the electrical bias (VDD) and optical bias (input laser light at the \textbf{IN} port) are maintained.

\subsubsection{Cross-coupled Photodiode Drive}
In the configuration shown in Figure \ref{psram_latch_config2}(a), the through and drop ports of M1 (M2) are connected to the waveguides connected to photodiodes P3 (P1) and P4 (P2), respectively. The output ports of the MRR M1 (M2) drive photodiodes P3 and P4 (P1 and P2), forming a cross-coupled structure to retain the stored data. The node Q (QB) drives the MRR M2 (M1). When Q = 1, M2 resonates with the incoming light, directing most of it to P2, which generates a higher current and forms a low-resistance path to GND, keeping QB at 0. In turn, QB keeps M1 out of resonance, allowing light to pass through to P3, which maintains Q at VDD. This cross-coupled MRRs and PDs structure reliably holds the stored data (Q = 1, and QB = 0), and the similar process applies when Q = 0 and QB = 1.

\subsubsection{Through-port only Cross-coupled Photodiode Drive}
In the configuration shown in Figure \ref{psram_latch_config2}(b), the through ports of MRRs M1 and M2 are used to drive the photodiodes (P1-P4) in a cross-coupled arrangement to latch the data. The through ports of M1 and M2 are split via optical 50:50 power splitters PS2 and PS3, directing the light to the photodiodes. The split through ports of M1 (M2) drive P2 and P3 (P1 and P4), forming a cross-coupled structure to retain the stored data. The node Q (QB) drives the MRR M2 (M1). As an example, when QB = 0, M1 is out of resonance, directing most of the light to P2 and P3, which generate higher currents, creating a low-resistance path to GND for QB and a low-resistance path to VDD for Q, keeping Q at 1 and QB at 0. In contrast, M2 is in resonance, and the incoming light to M2 is absorbed by the passive absorber (A2) connected to its drop port. Thus this cross-coupled MRRs and PDs structure reliably holds the stored data (Q = 1, QB = 0), and the similar mechanism also applies when Q = 0 and QB = 1.

\subsubsection{Drop-port only Cross-coupled Photodiode Drive}
In the configuration shown in Figure \ref{psram_latch_config2}(c), the drop ports of MRRs M1 and M2 are used to drive photodiodes P1-P4 in a cross-coupled setup to latch data. The drop ports of M1 and M2 are split using 50:50 optical power splitters PS2 and PS3, directing light to the photodiodes. The split drop ports of M1 (M2) drive P1 and P4 (P2 and P3), forming a cross-coupled structure to store data. The node Q (QB) drives the MRR M2 (M1). For instance, when Q = 1 (VDD) and QB = 0 (GND), these values are maintained as long as the optical and electrical biases are present. As Q = 1 and QB = 0, M2 resonates, sending most of the light to P2 and P3, generating higher currents that establish a low-resistance path to GND for QB and to VDD for Q, keeping Q at 1 and QB at 0. Meanwhile, M1 is out of resonance, and the incoming light to M1 is absorbed by the passive absorber (A1) at its through port. This cross-coupled MRRs and photodiodes structure reliably retains the stored data, with the similar process occurring when Q = 0 and QB = 1.

Among all the four configurations discussed above, the cross-coupled micro-ring resonator drive has fewer waveguide crossings compared to the cross-coupled photodiode drive, through-port only, or drop-port only photodiode drive configurations, resulting in the smallest area footprint. Additionally, the through-port only and drop-port only photodiode drive configurations require two extra power splitters and two passive absorbers, can lead to a larger area overhead. However, the through-port-only photodiode drive may provide a higher contrast between the on and off-resistance states of the photodiode path due to the greater variation in transmission between resonance conditions. Furthermore, both the through-port only and drop-port only photodiode drives may require higher bias laser power to compensate for losses introduced by the additional power splitters along the optical path. However, the availability of the drop and through ports allows for reading the stored data without requiring additional read MRRs, though at the cost of higher optical bias power through the IN port. Irrespective of the different cross-coupling structures, all these four configurations, which include micro-ring resonators (MRRs) M1 and M2 along with photodetectors (PDs) P1, P2, P3, and P4, constitute pSRAM latch structures capable of storing binary data at the storage nodes Q (data) and QB (complementary data). This data is retained as long as both the optical bias (laser input via \textbf{IN} port) and electrical bias (VDD) are maintained, similar to the operation of conventional electrical SRAM. For the remainder of this paper, we will focus on micro-ring resonator drive pSRAM latch configuration (discussed in the section \ref{sec_latch_mrr_drive}), although the write (discussed in section \ref{sec_write}) and read (discussed in section \ref{sec_read}) operations remain consistent regardless of the latch configuration.

\subsection{Write Operation of the photonic-SRAM (pSRAM) Bitcell} \label{sec_write}

Figure \ref{psram_write} illustrates the write structure for the pSRAM bitcell. To enable writing into the pSRAM bitcell, two input waveguides connected to the write bitline (\textbf{WBL}) and complementary write bitline (\textbf{WBLB}) ports are integrated with two additional 50:50 power splitters (PS2 and PS3) in the existing pSRAM structure as shown in the Figure \ref{psram_latch_config1}(a). Data is written by applying differential optical power through the \textbf{WBL} and \textbf{WBLB} connected waveguides. The \textbf{WBLB} (\textbf{WBL}) port is connected to photodiodes P1 and P4 (P2 and P3) via power splitters PS2 (PS3), facilitating differential write operations. To prevent write failures, higher optical write pulses are applied compared to the input optical bias (power from the \textbf{IN} port) to ensure the storage nodes Q and QB flip according to the new input data. 

\subsubsection{Differential Waveguide-based Driver-less Write Structure}\label{write_driverless}
\begin{figure}[!t]
\centering
\includegraphics[width=0.9\linewidth]{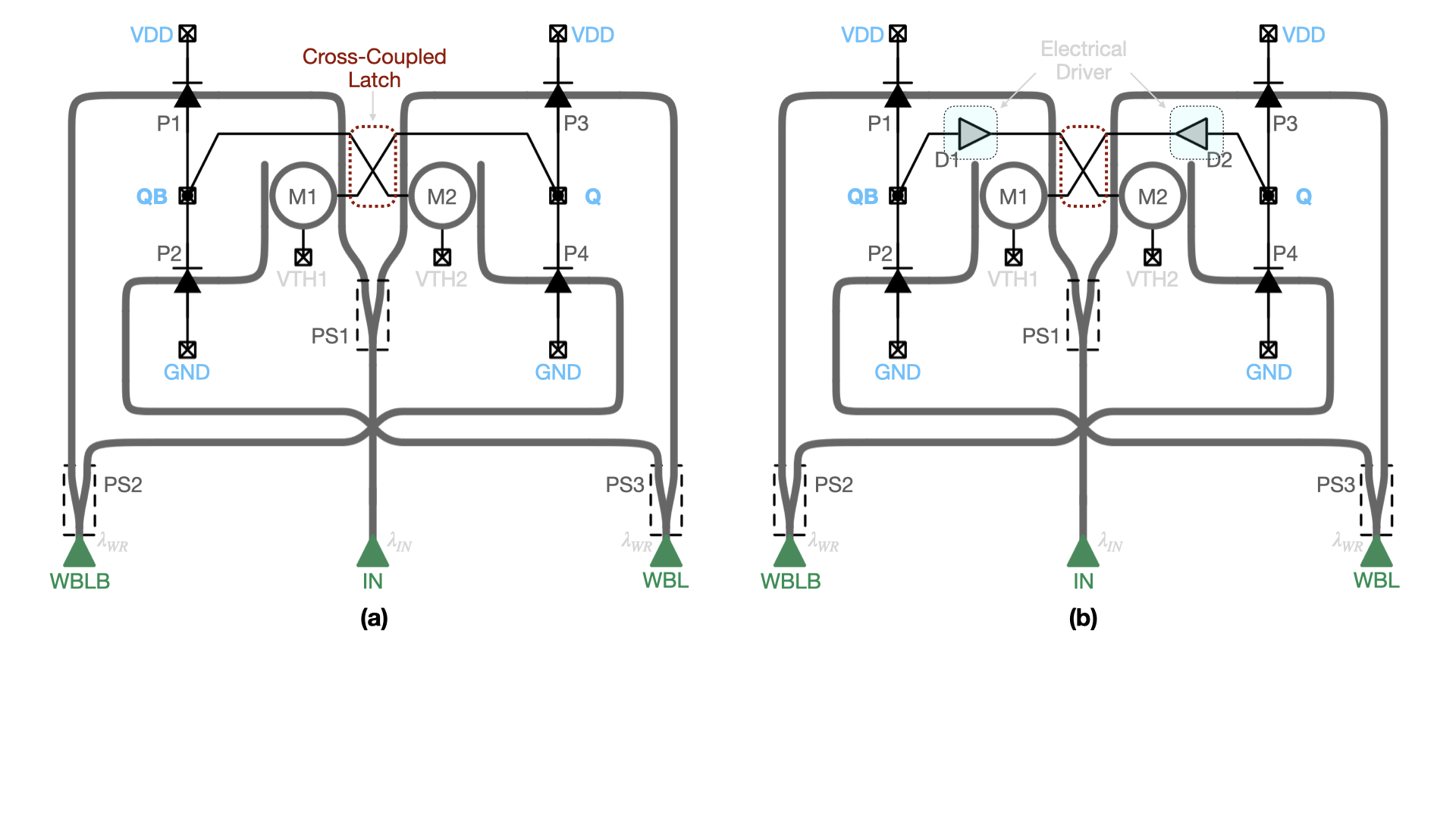}
\vspace{-20mm}
\caption{pSRAM write structure: (a) without electrical drivers, and (b) with electrical drivers, where D1 and D2 control MRR M2 and M1, respectively.}
\label{psram_write}
\end{figure}

Figure \ref{psram_write}(a) shows the structure without electrical drivers. To write new data (i.e., flip the stored data), differential optical power is applied to the \textbf{WBL} and \textbf{WBLB} connected waveguides. For instance, if the initial state is Q = 0 (GND) and QB = 1 (VDD), and the goal is to switch it to Q = 1 and QB = 0, a higher optical power is applied to the \textbf{WBL} connected waveguide, while no power is applied to the \textbf{WBLB} port. This results in P3 generating more current than P4, creating a low-resistance path to VDD for Q, raising Q to VDD and driving M1 into resonance. At the same time, QB drops to GND as P2 receives more light than P1, driving M2 out of resonance, thus stabilizing the state. Reversing the optical power between \textbf{WBL} and \textbf{WBLB} will flip the state to Q = 0 and QB = 1. The write speed is determined by the optical input power to the \textbf{WBL} and \textbf{WBLB} waveguides, with higher optical power leading to faster write speeds.

\subsubsection{Differential Waveguide-based Write Structure with Electrical Driver}\label{write_driver}
The write structure illustrated in Figure \ref{psram_write}(b) incorporates electrical drivers (cascaded inverter chain) to control the MRRs M1 and M2, enabling faster speeds but at the cost of additional electrical energy. The write mechanism follows the same principles as outlined in the previous section \ref{write_driverless}. A write optical pulse through the \textbf{WBL} sets Q = 1 and QB = 0, while a pulse through the \textbf{WBLB} sets Q = 0 and QB = 1 inside the pSRAM bitcell. The electrical drivers connected to nodes Q and QB increase the MRR driving current (as previous write structure as discussed in section \ref{write_driverless} utilizes resultant photodiode currents to drive the MRRs M1 and M2), facilitating high-speed operation.

\subsection{Read Operation of the photonic-SRAM (pSRAM) Bitcell} \label{sec_read}

Figure \ref{psram_read} illustrates the decoupled read structures of the pSRAM bitcell, comprising two micro-ring resonators (M3 and M4), an input read wordline waveguide connected to the \textbf{RWL} port, and differential output waveguides connected to the read bitline (\textbf{RBL}) and complementary read bitline (\textbf{RBLB}) ports. This configuration also incorporates a 50:50 power splitter (PS4) and two optical absorbers (A3 and A4) into the pSRAM bitcell design. The storage nodes Q and QB control the micro-ring resonators M4 and M3, respectively. During a read operation, a read laser activates the \textbf{RWL}, and the signals from the \textbf{RBL} and \textbf{RBLB} ports are processed at the periphery. The read MRRs M3 and M4 resonate with the optical input from the \textbf{RWL} when the voltage across them is VDD, and they are out of resonance when the voltage is GND. Additionally, the thermal ports VTH3 and VTH4 allow for the calibration of MRRs M3 and M4. This differential architecture ensures a reliable read mechanism, similar to its electrical counterparts.

\subsubsection{Decoupled Through-port Read Structure}\label{read_through}
\begin{figure}[!t]
\centering
\includegraphics[width=1\linewidth]{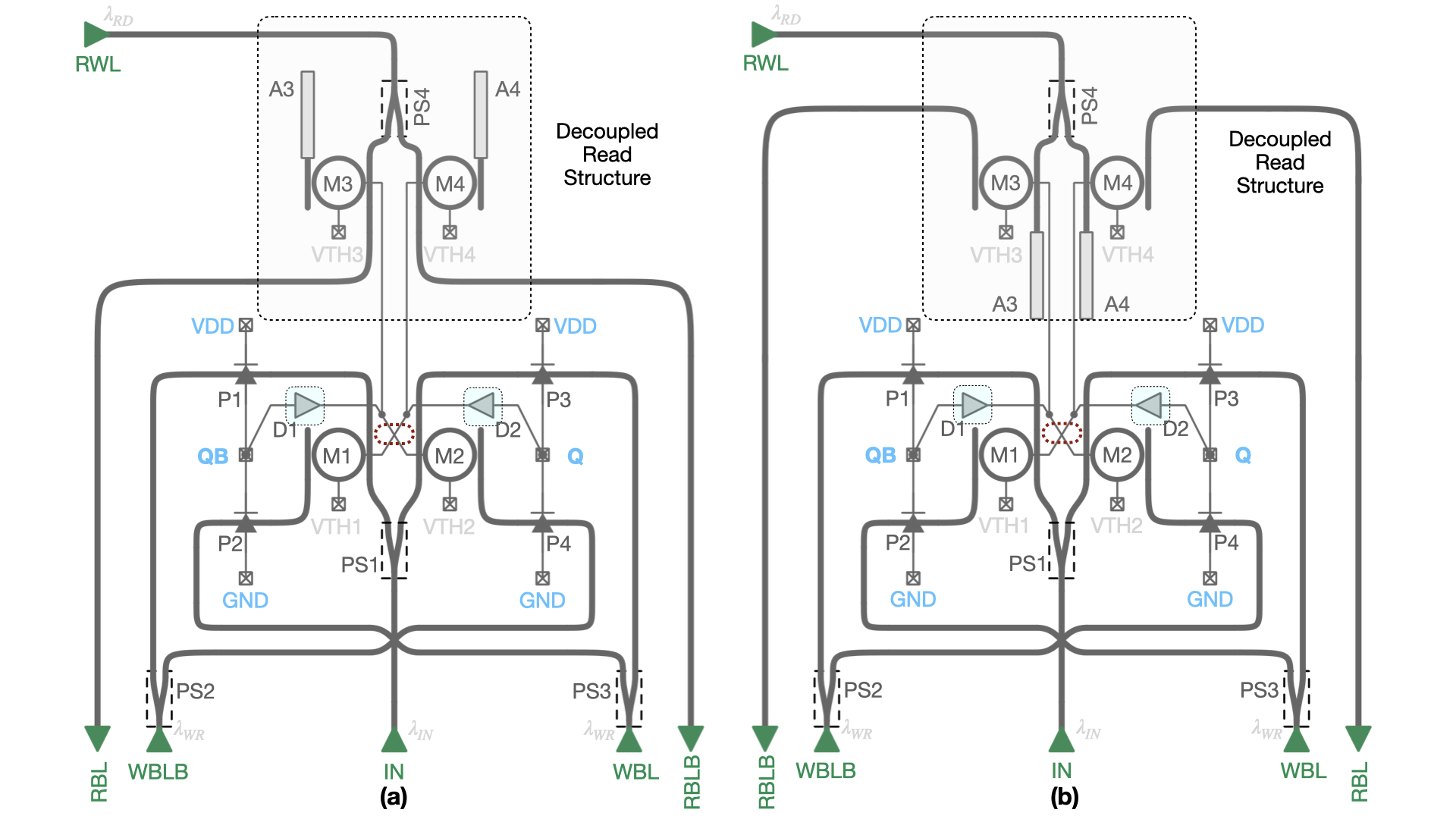}
\vspace{-5mm}
\caption{Decoupled Read Structures of pSRAM bitcell: (a) Through-Port Configuration and (b) Drop-Port Configuration.}
\label{psram_read}
\end{figure}

In this configuration (shown in Figure \ref{psram_read}(a)), the read bitline ports \textbf{RBL} and \textbf{RBLB} are connected to the through-ports of the read micro-ring resonators (MRRs) M3 and M4, respectively. The drop ports of M3 and M4 connect to the passive absorbers A3 and A4. To initiate a read operation on a bitcell, the read wordline (\textbf{RWL}) is activated. When Q = 1 (VDD), M4 enters resonance, preventing light from reaching the \textbf{RBLB}, while M3, which remains out of resonance since QB = 0 (GND), allows more light to pass through the \textbf{RBL} port. Conversely, in the state where Q = 0 and QB = 1, \textbf{RBLB} port receives a greater amount of light than \textbf{RBL}. An electro-optic sense amplifier then detects these optical signals and converts them into electrical data for subsequent processing in electrical domain. The decoupled read mechanism enhances stability during the read operation, preventing any read-disturb failures at the storage nodes. 

\subsubsection{Decoupled Drop-port Read Structure}\label{read_drop}
In this structure (illustrated in Figure \ref{psram_read}(b)), the read bitline ports \textbf{RBL} and \textbf{RBLB} are connected to the drop ports of the read micro-ring resonators (MRRs) M4 and M3, respectively. The through ports of M3 and M4 connect to the passive absorbers A3 and A4. To perform a read operation on a bitcell, the read wordline (RWL) is activated. When Q = 1 (VDD), M4 resonates, allowing light from the \textbf{RWL} to pass to the \textbf{RBL}, while M3 stays out of resonance due to QB = 0 (GND), leading to most light being absorbed by A3. In contrast, when Q = 0 and QB = 1, \textbf{RBLB} receives more light than \textbf{RBL} due to the in-resonance state of the read MRR M3. The decoupled read ports significantly enhance the robustness of the pSRAM structure against read-disturb failures. The thru-port read structure offers a greater sense margin between the on- and off-resonance states compared to the drop-port read structure; however, the drop-port configuration exhibits higher absolute power levels. It is important to note that these are not exhaustive variants, and alternative read structure designs could further optimize performance.

\section{Results and Discussions}

This section presents simulation results for the hold, write, and read operations of the proposed pSRAM bitcell. Beyond verifying nominal conditions, we simulate the photonic SRAM (pSRAM) under various critical scenarios—such as noise, laser wavelength shifts, process corners, and temperature variations—to ensure the bitcell's functional robustness. Furthermore, we analyze the bitcell's operating speed and both static and dynamic energy consumption, comparing our design's performance with previous works.

\subsection{Hold Operation Analysis}
In this subsection, we will examine the data retention stability of the pSRAM latch in response to noise and variations in laser wavelength.

\subsubsection{Transmission Spectra of MRR Optimization for Latching} 
\begin{figure}[!t]
\centering
\subfloat[]{\includegraphics[width=0.45\linewidth]{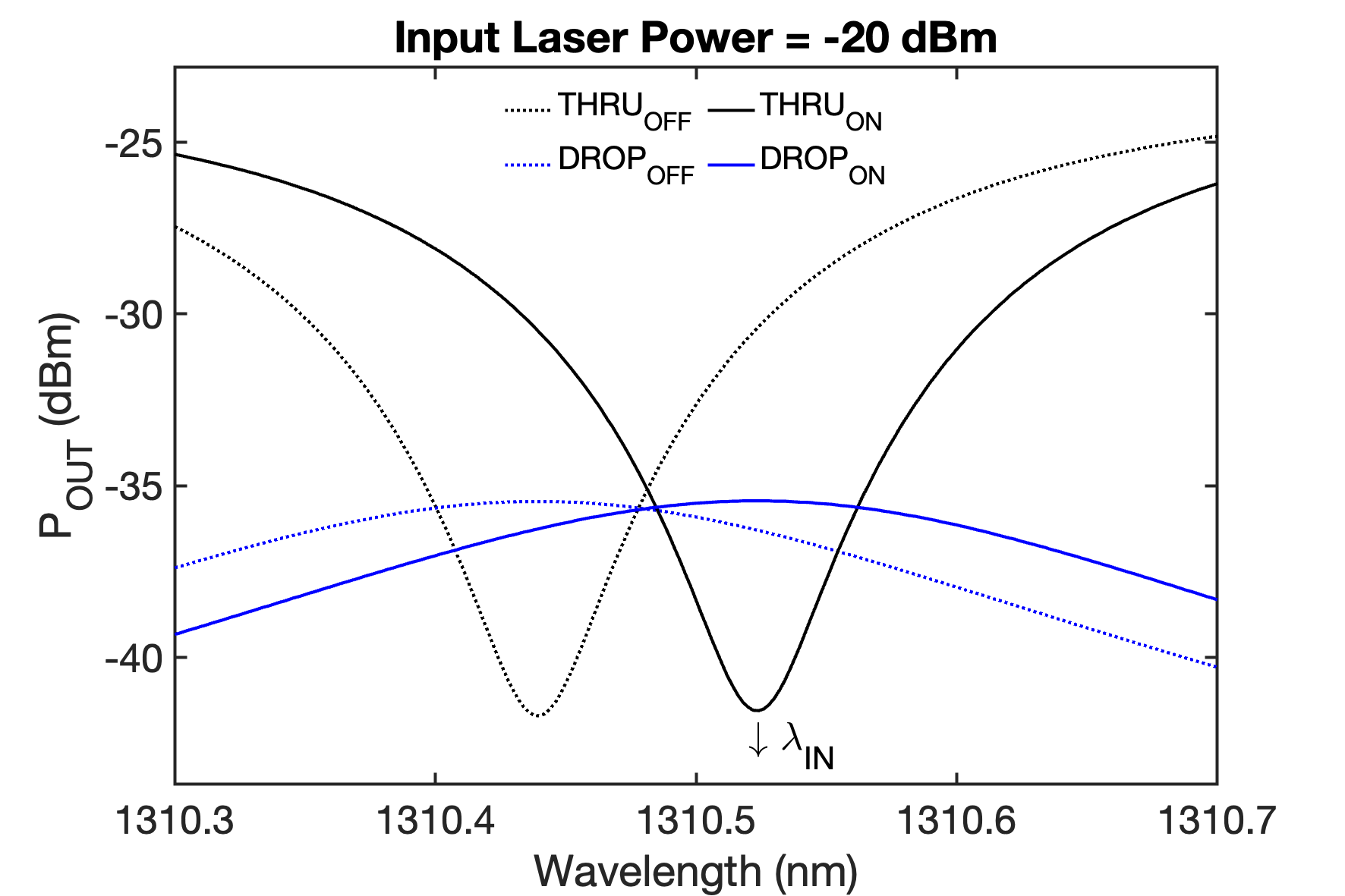}} 
\subfloat[]{\includegraphics[width=0.45\linewidth]{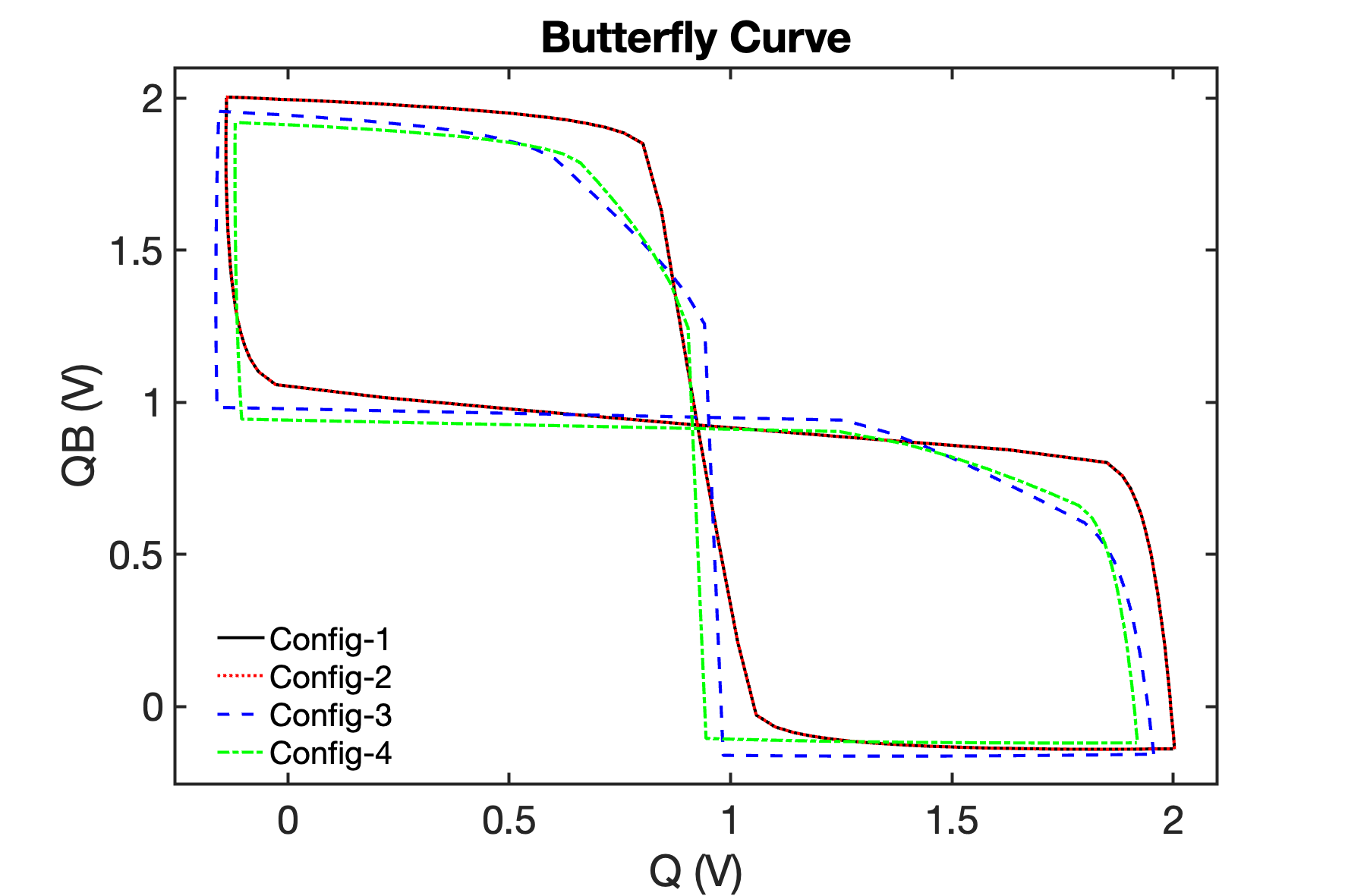}}
\caption{(a) Through-port and drop-port transmission spectra of the MRR for applied voltages of VDD and GND, and (b) Butterfly curves for four different photonic latch configurations showing hold static noise margin (HSNM).}
\label{mrr_char_hnm}
\end{figure}

Figure \ref{mrr_char_hnm}(a) presents the transmission spectra at the through-port and drop-port of the micro-ring resonator (MRR) for applied voltages of VDD and GND. The MRR has a ring radius of 7.5 \si{\mu}m, with gaps of 180 nm at the through-port and 395 nm at the drop-port, and a waveguide width of 400 nm used in the simulation. The coupling coefficient was chosen to maintain a minimum optical power difference of 3 dB between the through-port and drop-port in both the on-resonance and off-resonance states. When the voltage across the MRR's pn junction is VDD (GND), the MRR enters an on-resonance (off-resonance) state with the input laser wavelength, \si{\lambda_{IN}}. As illustrated in the figure, in the on-resonance state, the drop-port power (solid blue line) is 3 dB higher than the through-port power (solid black line) at \si{\lambda_{IN}}. In the off-resonance state, the through-port power (dotted black line) exceeds the drop-port power (dotted blue line) by 3 dB. This transmission characteristics ensures stable latch operation at the storage nodes within the balanced photodiode series structure.

\subsubsection{Hold Static Noise Margin (HSNM) and Butterfly Curve} 
Hold static noise margin (HSNM) is a key parameter that defines the maximum amount of static noise a cross-coupled latch can withstand without flipping the stored data. In SRAM, HSNM is critical for ensuring reliable data retention, especially during the hold state when no read or write operations are occurring. It essentially measures the robustness of the memory cell to external noise during this idle state. The stability of the hold state is typically estimated using a butterfly curve, which is generated by sweeping the voltage of one storage node (such as Q or QB) while simultaneously plotting the voltage of the opposing node (QB or Q) in the same graph. Due to the shape of the graph, it is often called the butterfly diagram. The largest square that can fit inside the two lobes of the curve represents the hold stability, with the minimum square size between the two lobes defining the HSNM. The wider the lobes, the more stable the latch, indicating a higher noise tolerance and stronger data retention capability. This analysis is crucial in ensuring that the memory cells can reliably maintain data integrity even in the presence of noise or disturbances.

Figure \ref{mrr_char_hnm}(b) shows the butterfly curves for four different photonic latch configurations: config-1: cross-coupled micro-ring resonator drive, config-2: cross-coupled photodiode drive, config-3: through-port only cross-coupled photodiode drive, and config-4: drop-port only cross-coupled photodiode drive. Configurations 1 and 2 exhibit slightly larger lobe areas, indicating better stability, compared to configurations 3 and 4, though all demonstrate large HSNM and robust bi-stable states, critical for effective latching. Considering the HSNM lobe area, along with the lower number of photonic components and waveguide crossings, configuration 1 has been used for further discussions, though any of these configurations could function as the core latch in a pSRAM bitcell.

\begin{figure}[!b]
\centering
\subfloat[]{\includegraphics[width=0.45\linewidth]{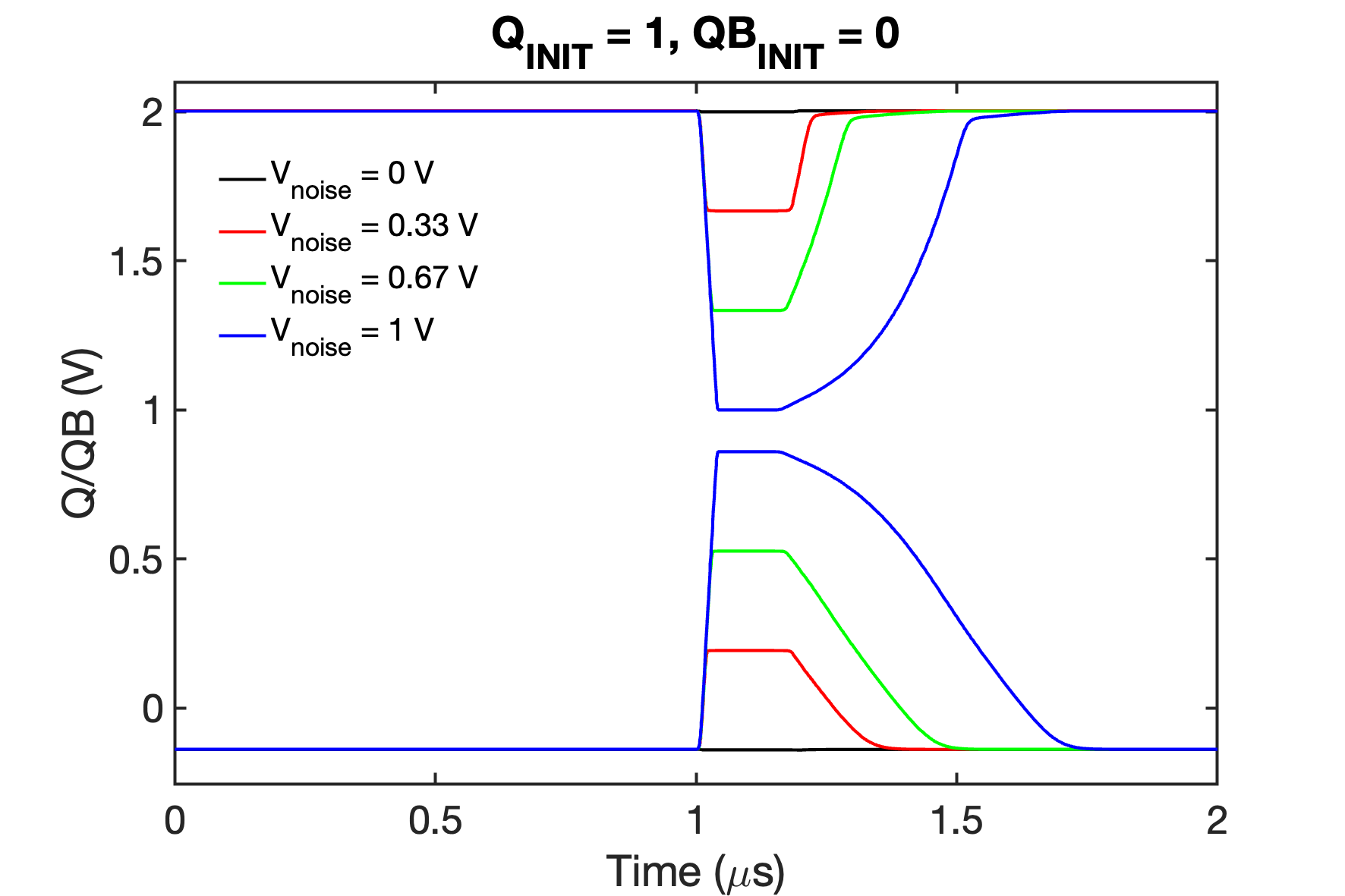}} 
\subfloat[]{\includegraphics[width=0.45\linewidth]{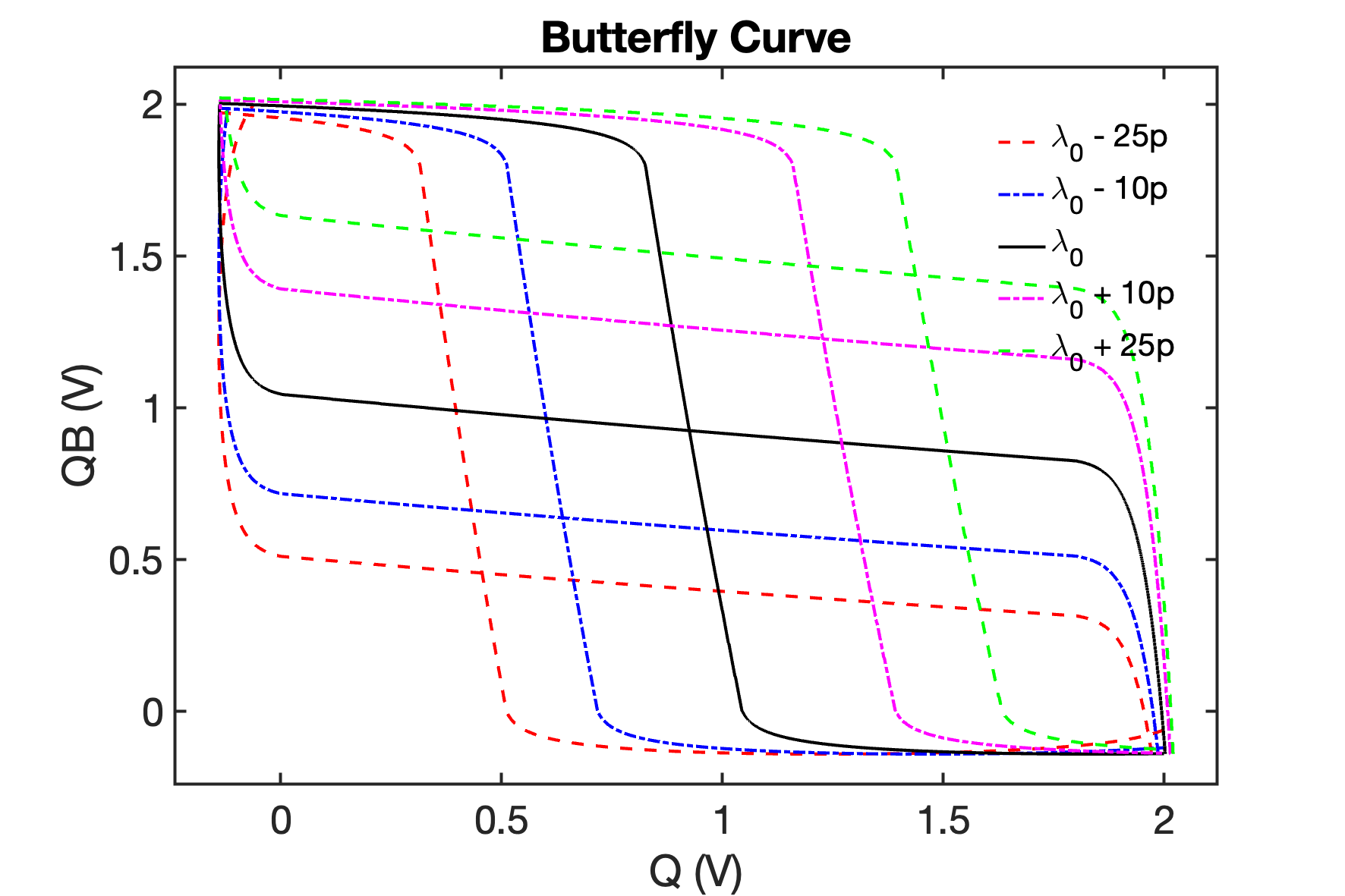}}
\caption{(a) Transient noise simulation for the pSRAM bitcell under differential voltage noise on the storage nodes, and (b) Butterfly curves showing the impact of input laser wavelength variation on the hold static noise margin (HSNM) for the pSRAM bitcell. Here, \si{\lambda_0} denotes the nominal wavelength at which the MRRs M1 and M2 resonate at an applied voltage of VDD.}
\label{hold_vnoise_wvnoise}
\end{figure}

\subsubsection{Differential Transient Voltage Noise} 
Figure \ref{hold_vnoise_wvnoise}(a) illustrates the transient noise simulation for the pSRAM bitcell (cross-coupled micro-ring drive). Initially, the stored values are Q = 1 and QB = 0, and the simulation runs for 2 \si{\mu}s. At the 1 \si{\mu}s, a differential transient noise pulse of 100 ns is introduced to both Q and QB nodes to simulate a worst-case scenario, attempting to flip the stored data. As shown in Figure \ref{hold_vnoise_wvnoise}(a), even with up to 1 V of noise on each Q and QB node (resulting in a 2 V differential between Q and QB), the pSRAM bitcell successfully regenerates the previously stored data. It is important to note that higher noise voltages increase the time required to restore the original state, but the bitcell continues to function correctly due to its regenerative behavior.

\subsubsection{Input Laser Wavelength Variation} 
The input optical laser wavelength can fluctuate due to electrical and thermal variations, environmental disturbances, and aging effects. However, these fluctuations can significantly affect the retention stability of photonic SRAM, as the latching depends on the optical resonance condition. A mismatch in wavelength can lead to a decrease in optical power output at both the through and drop ports, thereby reducing the hold static noise margin (HSNM). To demonstrate the robustness of the pSRAM, we simulated butterfly curves for various input wavelengths while keeping the micro-ring resonator (MRR) constant. Consequently, the output optical power transmission and the area of the hold noise margin lobes decreased. Within a range of approximately \textpm 25 pm, the bitcell can reliably retain data, displaying a robust bistable state (with two intersection points—one at VDD and the other at GND), with a reasonably large side lobe width (approximately 0.4 V in the worst case). Furthermore, static wavelength variations due to process or local mismatch can also be calibrated by modulating the resonance spectra of the MRR using its thermal port.

The pSRAM bitcell structure demonstrates a large hold static noise margin, making it resilient to reasonably high transient noise on the storage nodes and variations in the input laser wavelength. Overall, this configurations prove to be a suitable structure for a latch in the photonic bitcell architecture and exhibit enhanced data stability and reliability.

\subsection{Write Operation Analysis}

Figure \ref{psram_write_op} shows the simulation results of the write operation into the pSRAM bitcell. In the simulation, Q is initially set to 1 and QB to 0 (before 2 ns) to verify the data flip in the subsequent steps. To validate the write functionality, the data is first written as Q = 0 and QB = 1 (opposite to the initial state), followed by writing the opposite data, Q = 1 and QB = 0. The successful write operation is confirmed by observing the flipping of the storage nodes (Q and QB) in the simulated waveforms.

\subsubsection{Driver-less Write}
\begin{figure}[!b]
\centering
\subfloat[]{\includegraphics[width=0.45\linewidth]{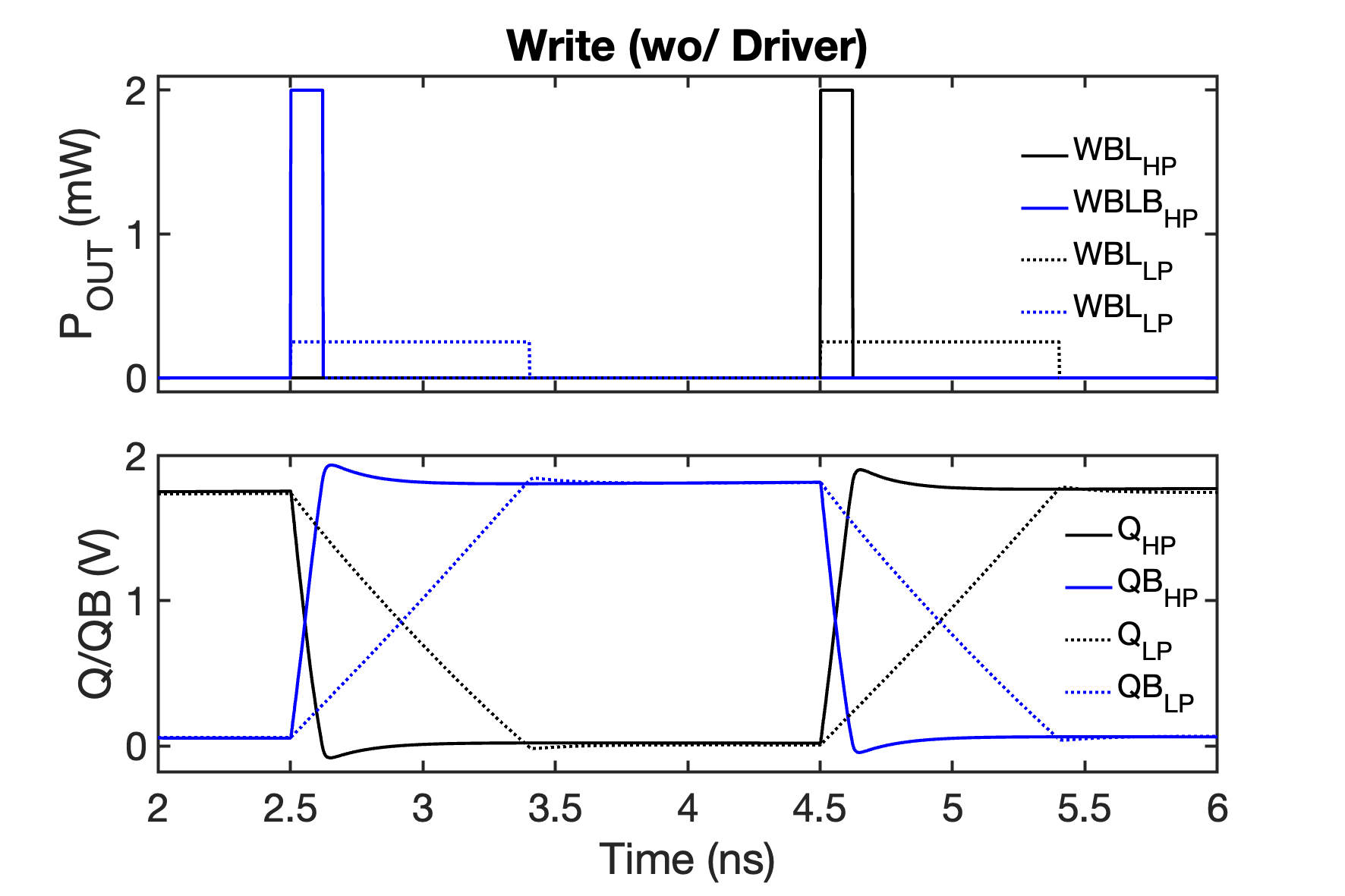}} 
\subfloat[]{\includegraphics[width=0.45\linewidth]{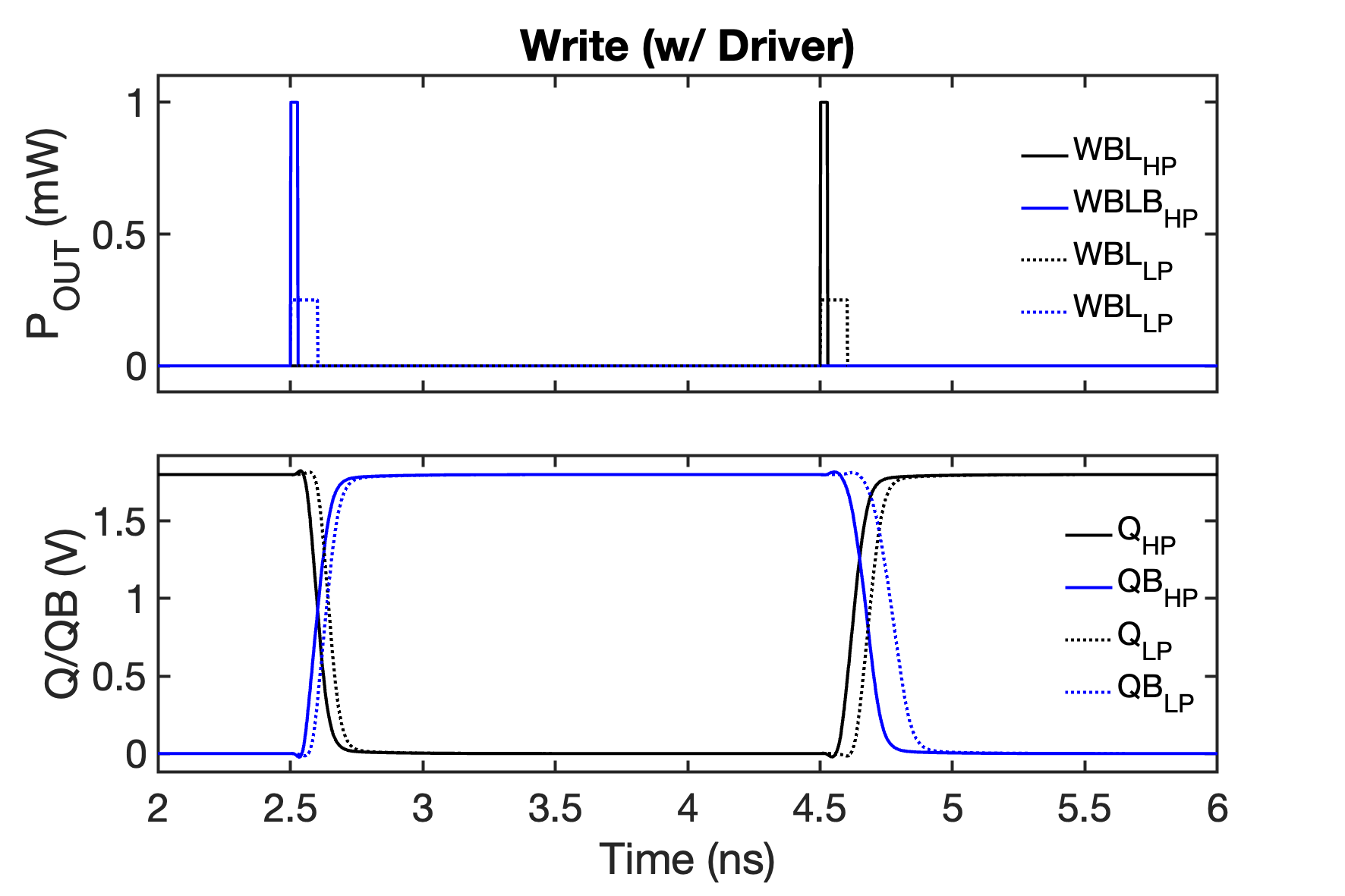}}
\caption{Simulation results of the write operation for the pSRAM bitcell utilizing (a) driver-less configuration, and (b) configuration with electrical drivers.}
\label{psram_write_op}
\end{figure}

Figure \ref{psram_write_op}(a) presents the write verification results of the pSRAM structure without electrical drivers, as discussed in section \ref{write_driverless}. In this configuration, the MRRs M1 and M2 are directly driven by the photocurrent generated by the balanced photodiodes. A successful write operation (storage node flipping) depends on the optical pulse intensity and pulse duration sent through the write bitline ports. High-speed write operations require high-intensity optical pulses, which generate higher resultant currents in the photodiode branches, driving the MRRs into on-resonance or off-resonance states depending on the optical power applied through the \textbf{WBL} and \textbf{WBLB} ports. The simulation results in Figure \ref{psram_write_op}(a) show two scenarios: high optical pulse intensity with short pulse width (2 mW, 120 ps) denoted as (WBL, WBLB, Q, QB)\si{_{HP}} and low optical pulse intensity with longer pulse width (0.25 mW, 900 ps) denoted as (WBL, WBLB, Q, QB)\si{_{LP}}. At 2.5 ns, the optical power is sent through the \textbf{WBLB} port (solid and dotted blue lines in the top subplot), causing the storage nodes to flip—Q changes from 1 to 0 (solid and dotted black lines in the bottom subplot) and QB changes from 0 to 1 (solid and dotted blue lines in the bottom subplot). This flip is driven by the optical pulse intensity and duration, which push MRRs M1 and M2 into opposite resonance states. Similarly, at 4.5 ns, optical power is sent through the \textbf{WBL} port (solid and dotted black lines in the top subplot), flipping the storage nodes again—Q changes from 0 to 1 (solid and dotted black lines in the bottom subplot) and QB from 1 to 0 (solid and dotted blue lines in the bottom subplot). 

\subsubsection{Write with Electrical Driver} \label{write_speed}
Figure \ref{psram_write_op}(b) illustrates the write verification waveforms for the pSRAM structure equipped with electrical drivers, as described in section \ref{write_driver}. In this structure, the micro-ring resonators M1 and M2 are driven by electrical drivers D1 and D2, as depicted in Figure \ref{psram_write}. The effectiveness of the write operation, which involves flipping the storage nodes, relies on the intensity and duration of the optical pulses transmitted through the \textbf{WBL} and \textbf{WBLB} ports. The simulation results presented in Figure \ref{psram_write_op}(b) encompass two scenarios: one featuring a high optical pulse intensity with a short pulse width (1 mW, 25 ps), labeled as (WBL, WBLB, Q, QB)\si{{HP}}, and the other showing a low optical pulse intensity with a longer pulse width (0.25 mW, 100 ps), labeled as (WBL, WBLB, Q, QB)\si{{LP}}. At 2.5 ns, optical power is directed through the \textbf{WBLB} port (indicated by the solid and dotted blue lines in the upper subplot), resulting in the storage nodes flipping—Q transitions from 1 to 0 (solid and dotted black lines in the lower subplot), while QB shifts from 0 to 1 (solid and dotted blue lines in the lower subplot). This transition is driven by the optical pulse's intensity and duration, which induce the micro-ring resonators M1 and M2 to enter opposite resonance states. Likewise, at 4.5 ns, optical power is transmitted through the \textbf{WBL} port (solid and dotted black lines in the upper subplot), leading to another flip of the storage nodes—Q changes from 0 to 1 (solid and dotted black lines in the lower subplot), and QB shifts from 1 to 0 (solid and dotted blue lines in the lower subplot). 

There is a trade-off between optical power and write speed; increased optical power leads to faster write operations. This is particularly noticeable in the high-optical pulse (HP) scenario, where Q and QB flip and settle faster than in the low optical pulse (LP) scenario for both structures (with and without an electrical driver). It is important to note that the maximum write speed of \textbf{40 GHz} can be achieved with the structure that includes an electrical driver with an optical pulse of 1 mW. While the electrical driver consumes additional energy, it enables faster operation of the micro-ring resonators compared to the direct driving method using the resultant current from photodiodes in the driver-less configuration. Moreover, the electrical driver enhances robustness and allows for quicker charging and discharging currents, contributing to improved speed.

\subsection{Read Operation Analysis}
Figure \ref{psram_read_op} presents the simulation results for the read operation of the pSRAM bitcell. In the waveforms, data is written into the bitcell and subsequently read through the read bitline ports (as shown in Figure \ref{psram_read}). To confirm the accuracy of the read functionality, the sequence begins by writing Q = 0 and QB = 1, followed by reading the data, then writing Q = 1 and QB = 0, and reading the data again. The successful read operation is validated by observing the optical power outputs at the read bitline ports (\textbf{RBL} and \textbf{RBLB}) during the read phase in the simulated waveforms.

\subsubsection{Through-port Read}\label{sim_read_through}
\begin{figure}[!t]
\centering
\subfloat[]{\includegraphics[width=0.6\linewidth]{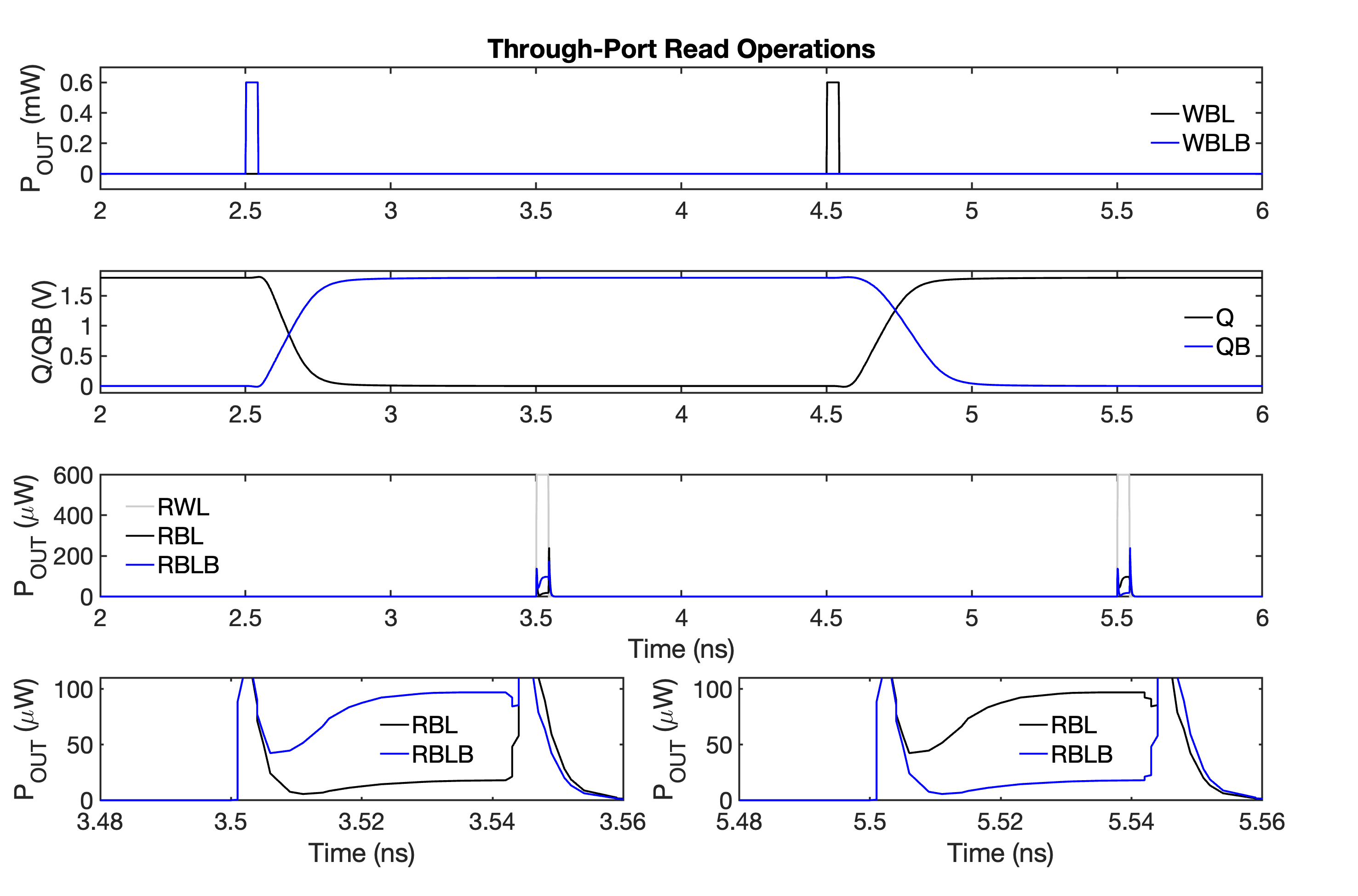}} \\
\subfloat[]{\includegraphics[width=0.6\linewidth]{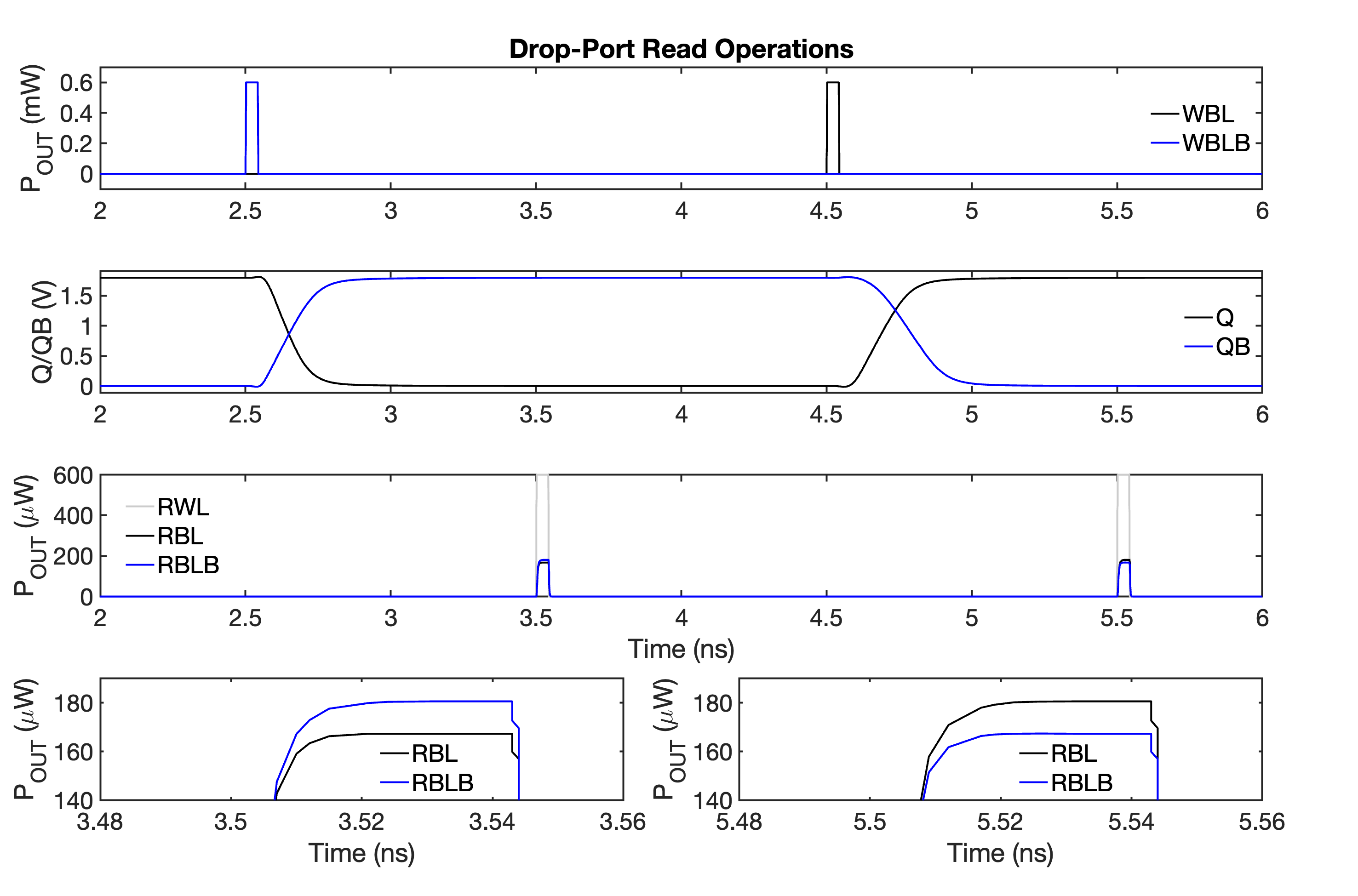}}
\caption{Simulation results of the read operation for the pSRAM bitcell utilizing (a) through-port read configuration, and (b) drop-port read configuration.}
\label{psram_read_op}
\end{figure}

Figure \ref{psram_read_op}(a) illustrates the read operation validation for the pSRAM structure, where the through-ports of the read micro-ring resonators (MRRs) M3 and M4 are connected to the read bitline ports \textbf{RBL} and \textbf{RBLB}, respectively, as detailed in section \ref{read_through}. The top two subplots display the correctness of the data written into the pSRAM bitcell. In the second subplot, data is written as Q = 0 and QB = 1 using an optical pulse through the \textbf{WBLB} port at 2.5 ns, and the opposite data, Q = 1 and QB = 0, is written via the \textbf{WBL} port at 4.5 ns. For the write operation, a 0.6 mW, 40 ps optical pulse is used to balance optical power and write speed, and a similar pulse duration is applied for the read process. The read wordline is activated through the \textbf{RWL} port at 3.5 ns and 5.5 ns to read data when Q = 0 and Q = 1, respectively. The bottom subplot zooms in on the read phase, showing that at 3.5 ns, the output at the \textbf{RBLB} port is higher, reflecting Q = 0 (QB = 1), and at 5.5 ns, the \textbf{RBL} port output is higher, corresponding to Q = 1 (QB = 1).

\subsubsection{Drop-port Read}
Figure \ref{psram_read_op}(b) illustrates the read operation for the pSRAM structure, where the drop ports of the read micro-ring resonators M3 and M4 are connected to the read bitline ports, \textbf{RBLB} and \textbf{RBL}, respectively, as outlined in section \ref{read_drop}. The same optical pulse parameters and verification methodology discussed in the previous section, \ref{sim_read_through}, are employed for the read verification in this case as well. The read wordline is activated with a 40 ps optical pulse through the \textbf{RWL} port at two designated times: 3.5 ns and 5.5 ns, enabling the retrieval of stored data values for Q = 0 and Q = 1, respectively. A zoomed-in view in the bottom subplot verifies the results of the read phase: at 3.5 ns, the \textbf{RBLB} port outputs a larger signal compared to the \textbf{RBL} port, indicating Q = 0 (with QB = 1), whereas at 5.5 ns, the \textbf{RBL} port generates a higher output than the \textbf{RBLB} port, corresponding to Q = 1 (with QB = 0). This validates the successful read operation of the stored data from the pSRAM structure.

The differences in optical power output between the \textbf{RBL} and \textbf{RBLB} ports are influenced by the extinction ratio of the read micro-ring resonators (MRRs), which measures the optical power outputs during their on-resonance and off-resonance states. By employing a decoupled read structure, the read MRRs can be optimized independently of the latch MRRs, enhancing the output differences observed at the read bitline ports. Additionally, while the optical pulse duration used in the read operation is the same as that in the write operation, the read process can be faster than the write operation. This is because the read operation does not require electro-optic activation, whereas the write operation involves driving the MRR with the current from the photodiode or an electrical driver. Furthermore, our differential read scheme resembles the read operation of traditional electrical SRAM, providing robustness against common mode noise sources.

\subsection{Process Variation Analysis}

\begin{figure}[!t]
\centering
\subfloat[]{\includegraphics[width=0.5\linewidth]{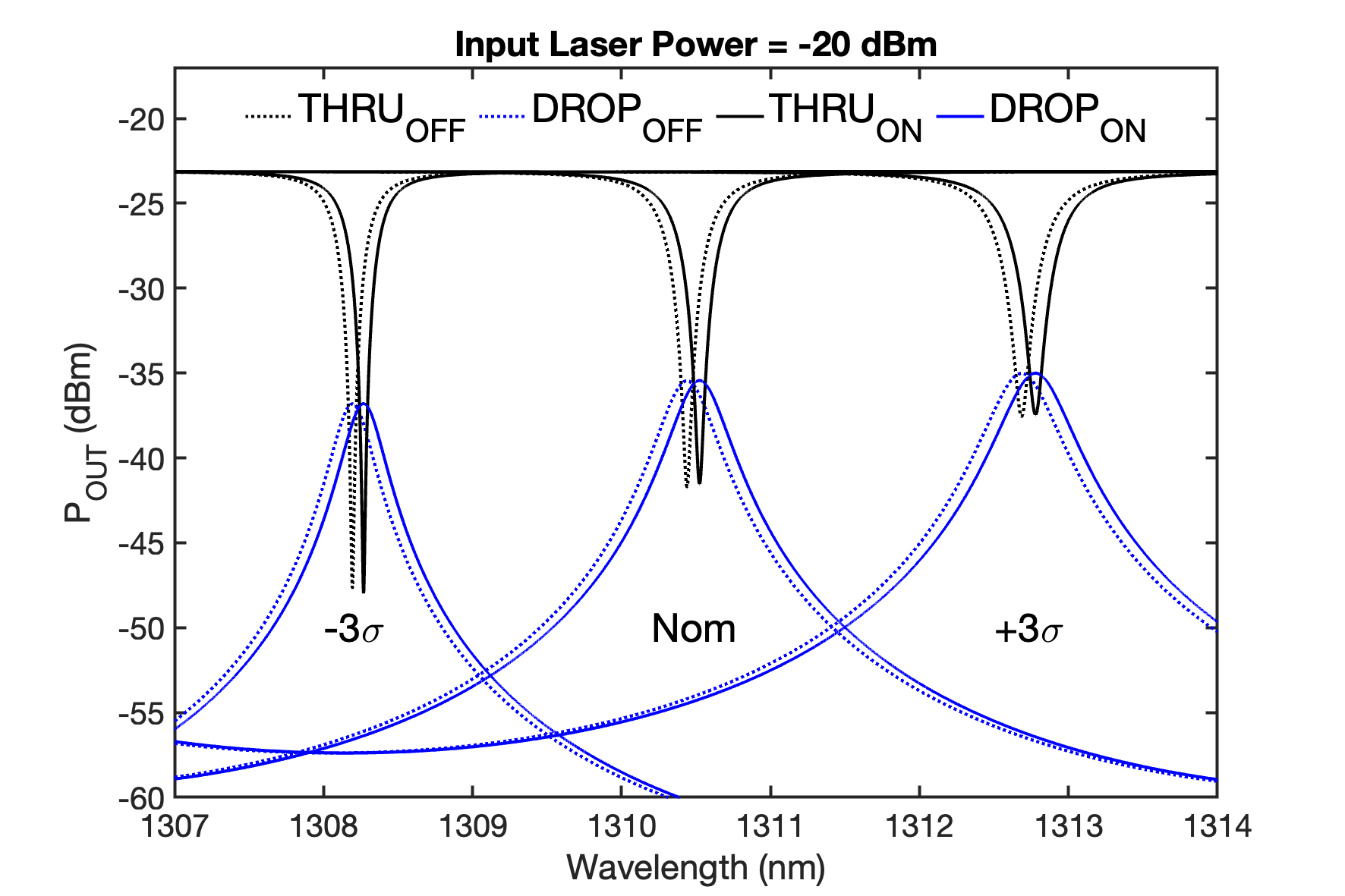}} \\
\subfloat[]{\includegraphics[width=0.5\linewidth]{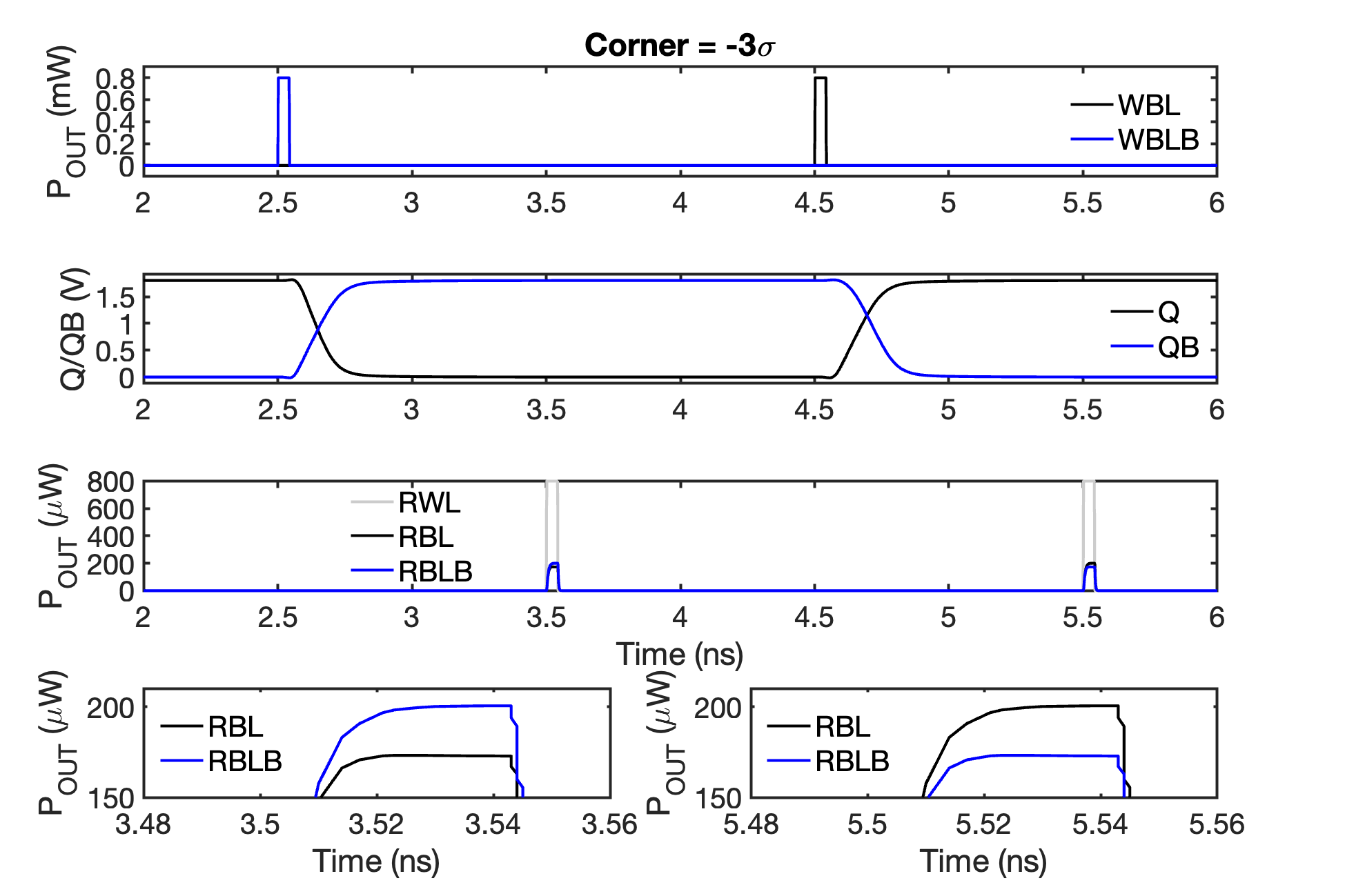}} \\
\subfloat[]{\includegraphics[width=0.5\linewidth]{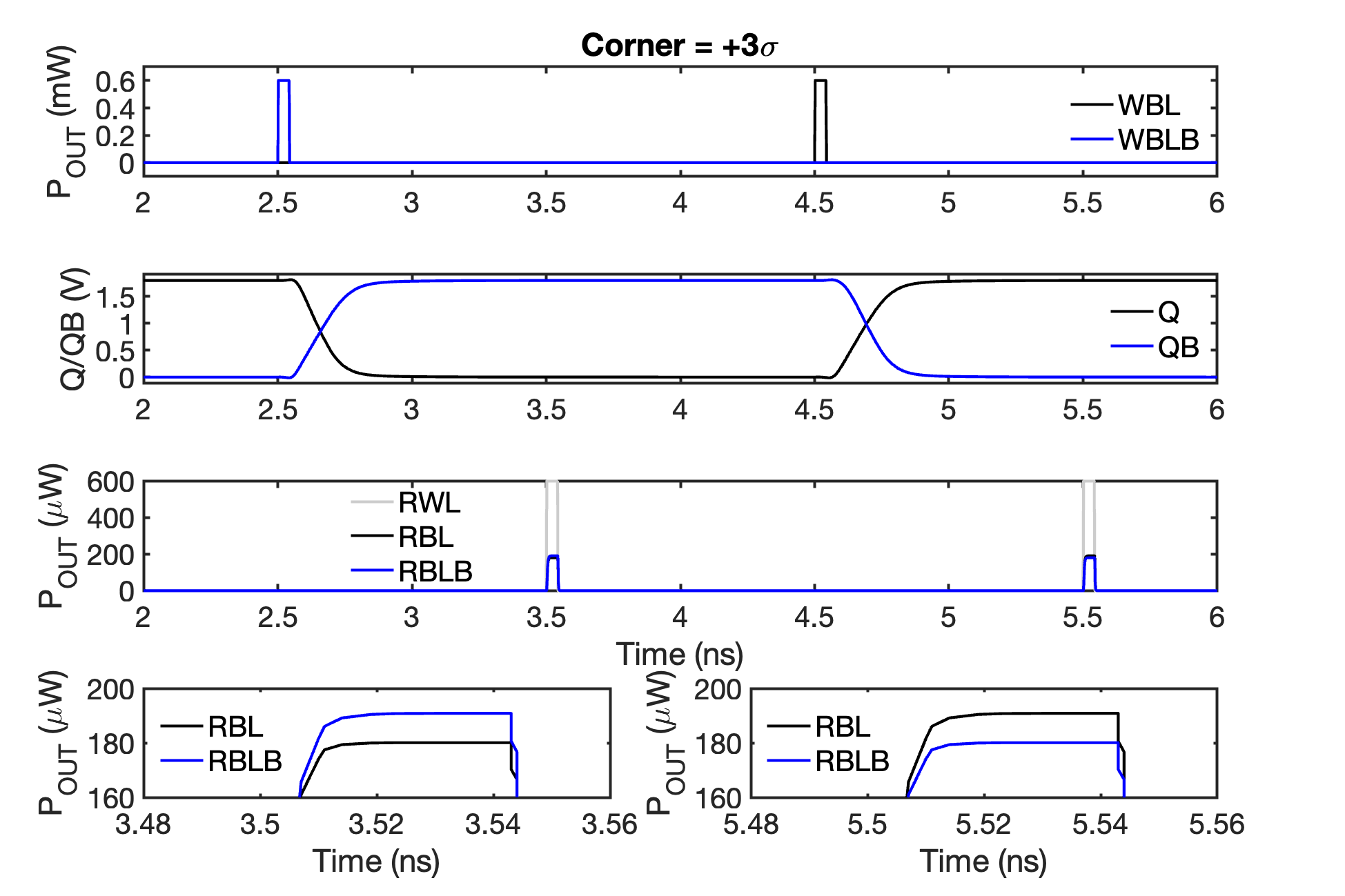}}
\caption{Process Variation Impact on (a) MRR Transmission Spectra and pSRAM write, hold and read verification at (b) -3\si{\sigma}, and (c) +3\si{\sigma} utilizing GF45SPCLO PDK.}
\label{psram_corner}
\end{figure}

Process variation significantly impacts the functionality of our proposed pSRAM bitcell, affecting both reliability and yield in large-scale fabrication. Such variations can lead to shifts in the resonance frequency of micro-ring resonators (MRRs), which may affect the pSRAM bitcell's data retention capabilities. Therefore, it is crucial to verify the functionality of our pSRAM cell across different process corners. Although each MRR features a thermal phase shifter port for tuning, managing multiple independent control ports in a large memory array may be impractical. A key advantage of our design is that if resonance peaks shift due to process variation, we can adjust the input source wavelength to align with the new resonance of the MRR, thus avoiding the need for additional thermal adjustments. Hence, we have validated the functionality of our pSRAM bitcell by accounting for \textpm3\si{\sigma} variation using the GF45SPCLO PDK by only adjusting the wavelength of the optical bias input through the \textbf{IN} port and the read wordline input through the \textbf{RWL} port to align with the MRRs' shifted resonance caused by process variation.

Figure \ref{psram_corner}(a) compares the transmission spectra of the MRR's through and drop ports under process variation (\textpm3\si{\sigma}) with the nominal case (typical corner). A blue shift is observed at -3\si{\sigma}, while a red shift occurs at +3\si{\sigma}. The figure also shows changes in optical output power levels with process variation. Notably, the power difference between the through-port and drop-port is greater at -3\si{\sigma} than at +3\si{\sigma}. While the MRR's coupling coefficients are optimized for the typical corner, the transmission spectra under these variations still provide sufficient hold margin to retain data.

Figure \ref{psram_corner}(b) shows the results of the write, hold, and read operations under a -3\si{\sigma} process variation. The top subplot depicts the application of the write pulse (0.8 mW, 40 ps) through the \textbf{WBLB} and \textbf{WBL} ports to store data for Q = 0 and Q = 1, respectively. The second subplot confirms successful data writing at the storage nodes (Q and QB), with data flipping at 2.5 ns and 4.5 ns. Data hold stability is demonstrated following the write phase. In the bottom zoomed subplot, the optical power during the read phase is higher at the \textbf{RBLB} port for Q = 0 and at the \textbf{RBL} port for Q = 1. The simulation adjusts only the source (\textbf{IN} port) and read wordline laser (\textbf{RWL}) wavelengths to account for the blue shift from 1310.5 nm to 1308.3 nm. Similarly, for a +3\si{\sigma} process variation, adjusting the source wavelength to 1312.8 nm enables successful data write and read operations, as shown in Figure \ref{psram_corner}(c).

\begin{figure}[!t]
\centering
\subfloat[]{\includegraphics[width=0.5\linewidth]{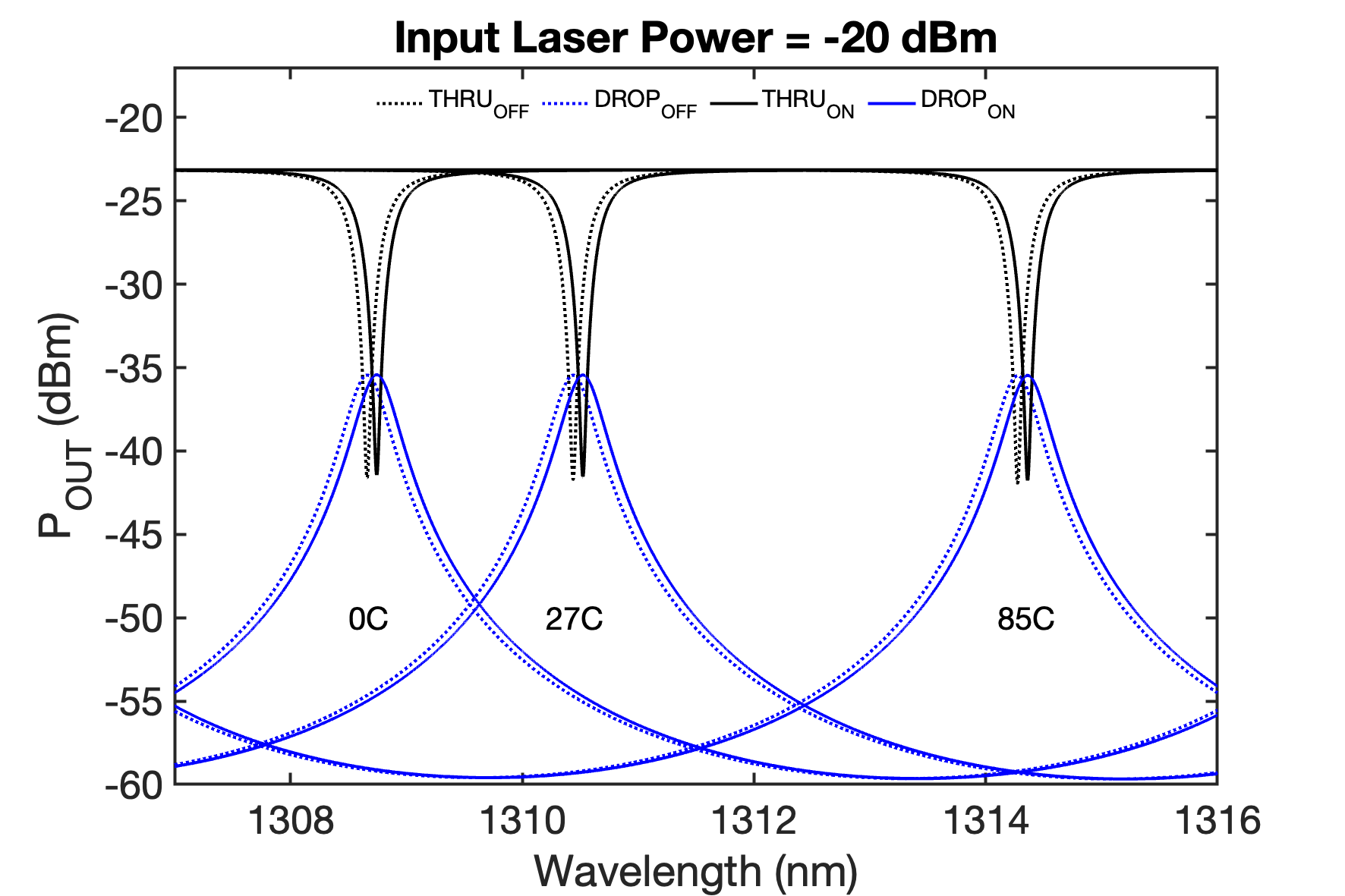}} \\
\subfloat[]{\includegraphics[width=0.5\linewidth]{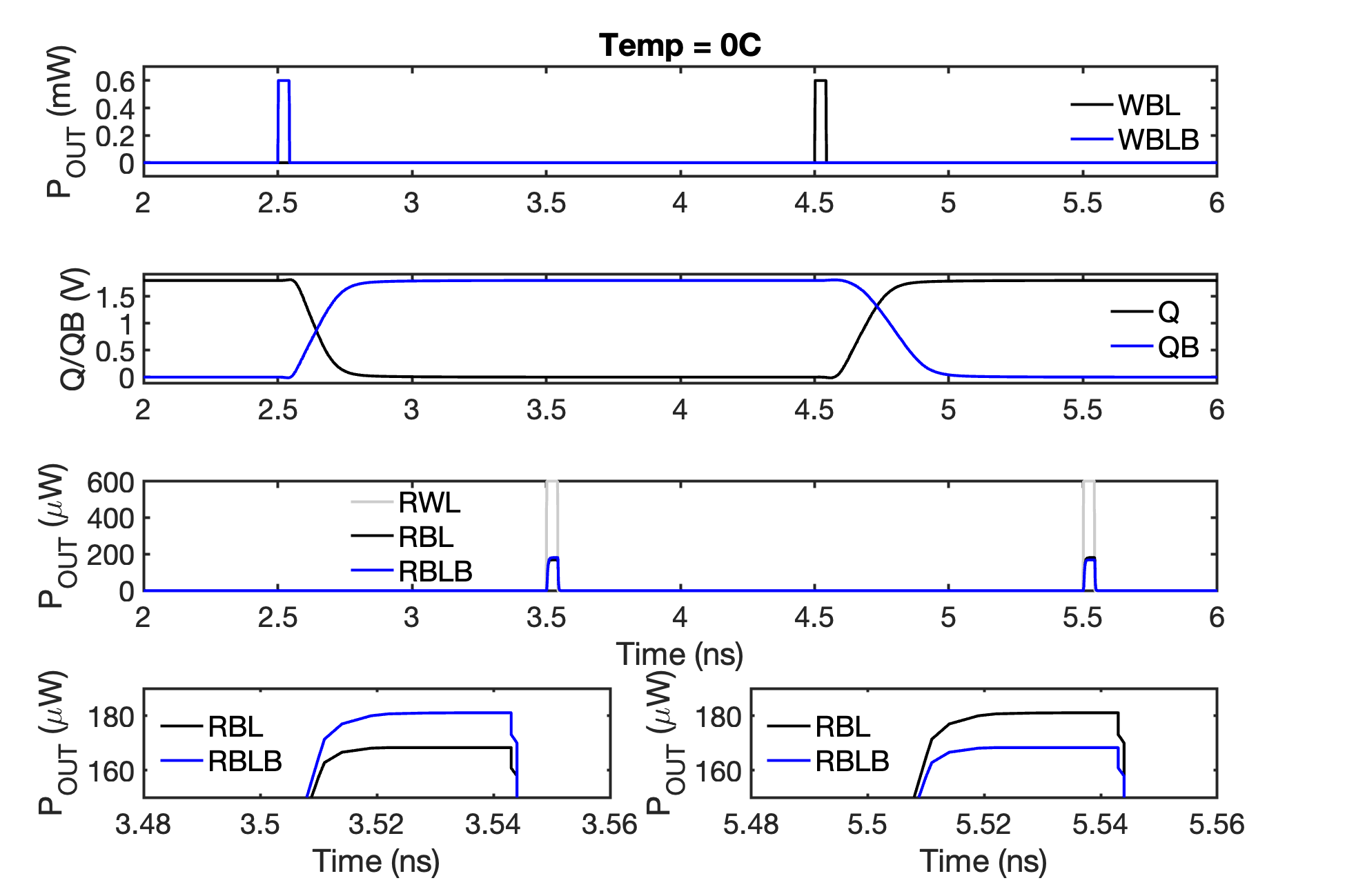}} \\
\subfloat[]{\includegraphics[width=0.5\linewidth]{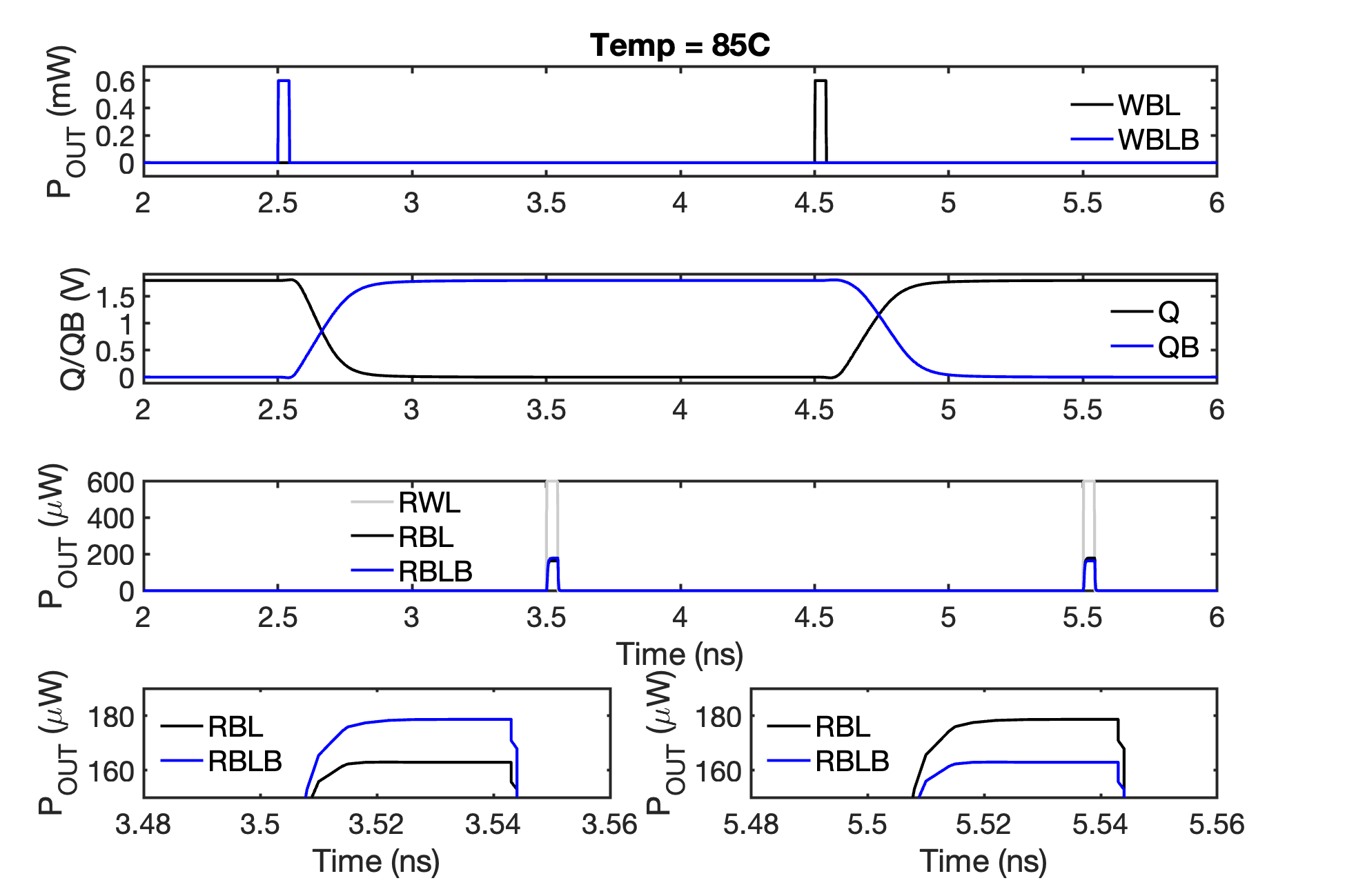}}
\caption{Temperature Variation Impact on (a) MRR Transmission Spectra and pSRAM write, hold and read verification at (b) 0\si{^\circ}C, and (c) 85\si{^\circ}C utilizing GF45SPCLO PDK.}
\label{psram_temp}
\end{figure}

\subsection{Temperature Variation Analysis}

Thermal variation can significantly affect the functionality of the proposed pSRAM bitcell by shifting the resonance peaks of the MRRs. A blue shift occurs at lower temperatures, and a red shift happens at higher temperatures, as depicted in Figure \ref{psram_temp}(a). Therefore, verifying the pSRAM bitcell's performance against temperature fluctuations is crucial to ensuring its robustness and reliability. Similar to calibrating against the process variation, adjusting the input source laser's wavelength (\textbf{IN} and \textbf{RWL}) to match the shifted resonance of the MRR allows the pSRAM to maintain reliable operation under temperature changes. By profiling the resonance shift with temperature, an on-chip temperature sensor could be integrated into an active loop with the source to automatically tune the wavelength. Additionally, the pSRAM operates with a low optical bias (10 \si{\mu}W), well below the self-heating threshold of the MRRs, ensuring minimal self-induced thermal effects. Thus, only adjustments to the source wavelength are required to maintain reliable pSRAM performance against environmental thermal variations.

Figure \ref{psram_temp}(b) and (c) show the results of the write, hold, and read operations at temperature 0\si{^\circ}C and 85\si{^\circ}C, respectively. The top subplot depicts the application of the write pulse (0.6 mW, 40 ps) through the \textbf{WBLB} and \textbf{WBL} ports to store data for Q = 0 and Q = 1, respectively at both temperature conditions. The second subplot confirms successful data writing at the storage nodes (Q and QB), with data flipping at 2.5 ns and 4.5 ns. Data retention stability is demonstrated following the write phase at both low and high temperature. In the bottom zoomed subplot, the optical power during the read phase is higher at the \textbf{RBLB} port for Q = 0 and at the \textbf{RBL} port for Q = 1. The simulation adjusts only the source (\textbf{IN} port) and read wordline laser (\textbf{RWL}) wavelengths to account for the blue and red shift at 0\si{^\circ}C and 85\si{^\circ}C temperature.

\subsection{Speed and Energy Analysis and Comparison with Existing Works}

As detailed in Section \ref{write_speed} and shown in Figure \ref{psram_write_op}(b), the maximum achievable speed for our proposed pSRAM bitcell is 40 GHz using a 1 mW, 25 ps optical write pulse with the pSRAM structure with the electrical driver. Notably, the addition of the electrical driver introduces an extra latency of approximately 140 ps (at typical corner and temperature of 27\si{^\circ}C) for the data transition during the write phase. Nevertheless, initiating the data flip (write) requires only a 25 ps duration with 1 mW optical power, allowing our pSRAM to operate at 40 GHz. Therefore, for memory-centric applications, our pSRAM bitcell can be accessed and utilized for computing at a speed of 40 GHz.  Although there is a power trade-off, higher speeds can be obtained at the expense of increased switching energy. The read speed is even faster, as no photodiode is involved, and the read micro-ring resonators are already settled by the electrical voltage node. Since no electrical activation is necessary for the read operation, the read speed is primarily limited by the speed of the peripheral opto-electronic conversion circuitry.

To determine the switching energy, we simulated the pSRAM bitcell using the GF45SPLCO PDK. The electrical energy was calculated by integrating the current over one phase and multiplying it by the supply voltage (VDD). Owing to the balanced photodiode (PD) structure, a static current flows; however, since the bias laser power in our work is 10 \si{\mu}W, and one photodiode in each branch receives minimal light, the resulting static current in the PD branch remains in the nA range. Moreover, the driver consumes energy to charge the capacitance of the latch and the read MRRs. The total electrical energy is combined with the optical write pulse energy, and we assume a 20\% wall-plug efficiency for the write laser. From our simulations, the total switching energy per operation is approximately 0.6 pJ. Additionally, we evaluated the static energy consumption of the bitcell, an important parameter for large memory arrays. The primary contributor is the static current in the PD branch, which is relatively small compared to the optical bias provided through the \textbf{IN} port. Considering the 10 \si{\mu}W laser input and a wall-plug efficiency of 0.2, the static energy consumption of the bitcell is around 0.03 pJ.




Table \ref{comparison} shows the comparison of performances of this work with various optical and electrical memory technologies reported in the literature. 

\begin{table}[!ht]
\caption{Performance comparison of various SRAMs.}
\label{comparison}
\begin{center}
\begin{tabular}{|c|c|c|c|c|c|c} \hline
\multirow{2}{*}{Device} & Speed  & Static Energy & Switching Energy  & Footprint & Technology\\ 
                        & (GHz) & (pJ/bit)      & (pJ/bit)          & \si{\mu m^2} &        \\ \hline
SOA-MZI \cite{SOA-MZI1, optical_ram_survey}               & 5  & 120 & 0.6 & 57.3\si{\times10^{3}} & Silicon coupled SOA-MZI\\
SOA-MZI \cite{SOA-MZI2, optical_ram_survey}               & 10 & 120  & 3 & 12\si{\times10^{6}} & Monolithic InP \\
Ring Laser \cite{mr_laser2, optical_ram_survey}           & 10 & 36 & 18 & 30\si{\times10^{3}} & Multiple Quantum Well (MQW)\\
Microdisk Laser$^*$ \cite{disk_laser, optical_ram_survey} & 10 & 0.6 & 0.002 & 56.25 & InP on SOI  \\
Nanocavity Laser$^*$ \cite{iii_v_nanocavity1}             & 10 & 0.01 & 0.0048 & 6.2 & III-V Photonic Crystal Nanocavity\\
PCM WGD \cite{pcm_ram1}                                   & 1 & --- & 13.4 & 0.16 & PCM on Silicon Nitride Waveguide \\ 
e-SRAM \cite{e_SRAM}                                      & 2.7 & --- & --- & 0.05 & Silicon-only \\
\textbf{This Work$^\dagger$}                                        & \textbf{40} & \textbf{0.03} & \textbf{0.6} & 95.7\si{\times10^{3}} & \textbf{Silicon-only} \\ 
\hline
\multicolumn{4}{l}{$^*$ \scriptsize{Excluding wavelength tuning energy consumption.}} \\
\multicolumn{4}{l}{$^\dagger$ \scriptsize{Wall-plug Efficiency = 0.2}} \\
\end{tabular}
\end{center}
\end{table}

\subsection{Potential Applications}

Our proposed photonic-SRAM (pSRAM) bitcell, while functionally equivalent to traditional electrical SRAM with true static behavior, takes advantage of silicon photonics for high-performance computing. Using fabrication-friendly components such as micro-ring resonators, photodiodes, and power splitters, the pSRAM seamlessly integrates with existing silicon photonics manufacturing processes and can also be monolithically fabricated with electrical PDKs (e.g. GlobalFoundries GF45SPCLO PDK). This compatibility allows for the creation of 2D pSRAM arrays, forming high-bandwidth memory (HBM) banks that support various configurations similar to electrical SRAM, such as flip-flop, first-in-first-out (FIFO), last-in-first-out (LIFO), and dual-port memory systems. Additionally, 2D arrays can facilitate cross-bar analog and digital computing in memory, as shown in prior studies \cite{photonic_imc_crossbar, photonic_bnn, photonic_imc_neuromorphic}. Using multiple wavelengths within a single waveguide, pSRAM enables photonic in-memory computing, allowing massive parallel matrix multiplication, such as in photonic tensor cores \cite{photonic_tensor_core}. This parallelism greatly enhances throughput, positioning it as a powerful solution for applications that demand high tera-operations per second (TOPS), outperforming electrical alternatives. Moreover, the design supports ultra-high-speed read/write operations for real-time data access and processing \cite{photonic_sp}. The structure can also be scaled down further by reducing the size of the micro-ring resonators, increasing memory density while preserving energy efficiency. Ultimately, the pSRAM architecture offers significant advancements in next-generation memory technologies, delivering ultra-fast speed, large bandwidth, and seamless integration with both photonic and electronic systems.

\section{Conclusion}

In summary, this work presents comprehensive design and validation of a high-speed and energy-efficient differential photonic SRAM (pSRAM) bitcell that utilizes cross-coupled micro-ring resonators and photodiodes, developed with GlobalFoundries 45nm CMOS-compatible Silicon Photonics (GF45CLO) platform. We validated the hold, read, and write functionalities of the pSRAM bitcell against various corner cases, temperature variations, and fluctuations in the input source's wavelength to ensure the structure's robustness and high yield for large-scale manufacturing. This architecture is not only fabrication-friendly, facilitating smooth integration with conventional electrical systems but also aligns well with current silicon photonics foundry processes, making it particularly suitable for large-scale volume production. Furthermore, the pSRAM shows significant promise for ultra-fast, scalable memory arrays due to its high speed, energy efficiency, and scalable design that can create new opportunities for photonic computing, especially in high-speed data processing and storage applications. This work lays the groundwork for next-generation memory technologies, advancing the integration of photonics into mainstream computing by merging the speed and bandwidth of optical systems with the scalability and reliability of silicon-based manufacturing.

\bibliography{sample}

\begin{thebibliography}{10}
\urlstyle{rm}
\expandafter\ifx\csname url\endcsname\relax
  \def\url#1{\texttt{#1}}\fi
\expandafter\ifx\csname urlprefix\endcsname\relax\def\urlprefix{URL }\fi
\expandafter\ifx\csname doiprefix\endcsname\relax\def\doiprefix{DOI: }\fi
\providecommand{\bibinfo}[2]{#2}
\providecommand{\eprint}[2][]{\url{#2}}

\bibitem{eSRAM_prb}
\bibinfo{author}{Cho, K.} \emph{et~al.}
\newblock \bibinfo{journal}{\bibinfo{title}{Sram write-and performance-assist cells for reducing interconnect resistance effects increased with technology scaling}}.
\newblock {\emph{\JournalTitle{IEEE Journal of Solid-State Circuits}}} \textbf{\bibinfo{volume}{57}}, \bibinfo{pages}{1039--1048} (\bibinfo{year}{2022}).

\bibitem{memory_wall_bottleneck}
\bibinfo{author}{Wulf, W.~A.} \& \bibinfo{author}{McKee, S.~A.}
\newblock \bibinfo{journal}{\bibinfo{title}{Hitting the memory wall: Implications of the obvious}}.
\newblock {\emph{\JournalTitle{ACM SIGARCH computer architecture news}}} \textbf{\bibinfo{volume}{23}}, \bibinfo{pages}{20--24} (\bibinfo{year}{1995}).

\bibitem{optics_advantage1}
\bibinfo{author}{Shaker, L.~M.}, \bibinfo{author}{Al-Amiery, A.}, \bibinfo{author}{Isahak, W. N. R.~W.} \& \bibinfo{author}{Al-Azzawi, W.~K.}
\newblock \bibinfo{journal}{\bibinfo{title}{Integrated photonics: bridging the gap between optics and electronics for enhancing information processing}}.
\newblock {\emph{\JournalTitle{Journal of Optics}}} \bibinfo{pages}{1--13} (\bibinfo{year}{2023}).

\bibitem{optics_advantage2}
\bibinfo{author}{Kibebe, C.~G.}, \bibinfo{author}{Liu, Y.} \& \bibinfo{author}{Tang, J.}
\newblock \bibinfo{journal}{\bibinfo{title}{Harnessing optical advantages in computing: a review of current and future trends}}.
\newblock {\emph{\JournalTitle{Frontiers in Physics}}} \textbf{\bibinfo{volume}{12}}, \bibinfo{pages}{1379051} (\bibinfo{year}{2024}).

\bibitem{optics_advantage3}
\bibinfo{author}{Maniotis, P.}, \bibinfo{author}{Fitsios, D.}, \bibinfo{author}{Kanellos, G.~T.} \& \bibinfo{author}{Pleros, N.}
\newblock \bibinfo{journal}{\bibinfo{title}{Optical buffering for chip multiprocessors: a 16ghz optical cache memory architecture}}.
\newblock {\emph{\JournalTitle{Journal of lightwave technology}}} \textbf{\bibinfo{volume}{31}}, \bibinfo{pages}{4175--4191} (\bibinfo{year}{2013}).

\bibitem{dl_ref}
\bibinfo{author}{LeCun, Y.}, \bibinfo{author}{Bengio, Y.} \& \bibinfo{author}{Hinton, G.}
\newblock \bibinfo{journal}{\bibinfo{title}{Deep learning}}.
\newblock {\emph{\JournalTitle{nature}}} \textbf{\bibinfo{volume}{521}}, \bibinfo{pages}{436--444} (\bibinfo{year}{2015}).

\bibitem{tpu_ref}
\bibinfo{author}{Jouppi, N.~P.} \emph{et~al.}
\newblock \bibinfo{title}{In-datacenter performance analysis of a tensor processing unit}.
\newblock In \emph{\bibinfo{booktitle}{Proceedings of the 44th annual international symposium on computer architecture}}, \bibinfo{pages}{1--12} (\bibinfo{year}{2017}).

\bibitem{optical_ram_survey}
\bibinfo{author}{Alexoudi, T.} \& \bibinfo{author}{et~al.}
\newblock \bibinfo{journal}{\bibinfo{title}{Optical ram and integrated optical memories: a survey}}.
\newblock {\emph{\JournalTitle{Light: Science \& Applications}}} \textbf{\bibinfo{volume}{9}}, \bibinfo{pages}{91} (\bibinfo{year}{2020}).

\bibitem{SOA-MZI1}
\bibinfo{author}{Pleros, N.}, \bibinfo{author}{Apostolopoulos, D.}, \bibinfo{author}{Petrantonakis, D.}, \bibinfo{author}{Stamatiadis, C.} \& \bibinfo{author}{Avramopoulos, H.}
\newblock \bibinfo{journal}{\bibinfo{title}{Optical static ram cell}}.
\newblock {\emph{\JournalTitle{IEEE Photonics Technology Letters}}} \textbf{\bibinfo{volume}{21}}, \bibinfo{pages}{73--75} (\bibinfo{year}{2008}).

\bibitem{SOA-MZI2}
\bibinfo{author}{Pitris, S.} \emph{et~al.}
\newblock \bibinfo{journal}{\bibinfo{title}{Wdm-enabled optical ram at 5 gb/s using a monolithic inp flip-flop chip}}.
\newblock {\emph{\JournalTitle{IEEE Photonics Journal}}} \textbf{\bibinfo{volume}{8}}, \bibinfo{pages}{1--7} (\bibinfo{year}{2016}).

\bibitem{SOA-XGM1}
\bibinfo{author}{Fitsios, D.}, \bibinfo{author}{Vagionas, C.}, \bibinfo{author}{Kanellos, G.~T.}, \bibinfo{author}{Miliou, A.} \& \bibinfo{author}{Pleros, N.}
\newblock \bibinfo{journal}{\bibinfo{title}{Dual-wavelength bit input optical ram with three soa-xgm switches}}.
\newblock {\emph{\JournalTitle{IEEE Photonics Technology Letters}}} \textbf{\bibinfo{volume}{24}}, \bibinfo{pages}{1142--1144} (\bibinfo{year}{2012}).

\bibitem{SOA-XGM2}
\bibinfo{author}{Vagionas, C.}, \bibinfo{author}{Fitsios, D.}, \bibinfo{author}{Kanellos, G.~T.}, \bibinfo{author}{Pleros, N.} \& \bibinfo{author}{Miliou, A.}
\newblock \bibinfo{journal}{\bibinfo{title}{Optical ram and flip-flops using bit-input wavelength diversity and soa-xgm switches}}.
\newblock {\emph{\JournalTitle{Journal of lightwave technology}}} \textbf{\bibinfo{volume}{30}}, \bibinfo{pages}{3003--3009} (\bibinfo{year}{2012}).

\bibitem{SOA-Ring_Laser}
\bibinfo{author}{Wang, J.}, \bibinfo{author}{Meloni, G.}, \bibinfo{author}{Berrettini, G.}, \bibinfo{author}{Poti, L.} \& \bibinfo{author}{Bogoni, A.}
\newblock \bibinfo{journal}{\bibinfo{title}{All-optical clocked flip-flops and binary counting operation using soa-based sr latch and logic gates}}.
\newblock {\emph{\JournalTitle{IEEE Journal of Selected Topics in Quantum Electronics}}} \textbf{\bibinfo{volume}{16}}, \bibinfo{pages}{1486--1494} (\bibinfo{year}{2010}).

\bibitem{mr_laser1}
\bibinfo{author}{Hill, M.~T.} \emph{et~al.}
\newblock \bibinfo{journal}{\bibinfo{title}{A fast low-power optical memory based on coupled micro-ring lasers}}.
\newblock {\emph{\JournalTitle{nature}}} \textbf{\bibinfo{volume}{432}}, \bibinfo{pages}{206--209} (\bibinfo{year}{2004}).

\bibitem{mr_laser2}
\bibinfo{author}{Trita, A.} \emph{et~al.}
\newblock \bibinfo{journal}{\bibinfo{title}{Monolithic all-optical set-reset flip-flop operating at 10 gb/s}}.
\newblock {\emph{\JournalTitle{IEEE Photonics Technology Letters}}} \textbf{\bibinfo{volume}{25}}, \bibinfo{pages}{2408--2411} (\bibinfo{year}{2013}).

\bibitem{mr_laser3}
\bibinfo{author}{Wang, Z.}, \bibinfo{author}{Yuan, G.}, \bibinfo{author}{Verschaffelt, G.}, \bibinfo{author}{Danckaert, J.} \& \bibinfo{author}{Yu, S.}
\newblock \bibinfo{journal}{\bibinfo{title}{Storing 2 bits of information in a novel single semiconductor microring laser memory cell}}.
\newblock {\emph{\JournalTitle{IEEE Photonics Technology Letters}}} \textbf{\bibinfo{volume}{20}}, \bibinfo{pages}{1228--1230} (\bibinfo{year}{2008}).

\bibitem{disk_laser}
\bibinfo{author}{Liu, L.} \emph{et~al.}
\newblock \bibinfo{journal}{\bibinfo{title}{An ultra-small, low-power, all-optical flip-flop memory on a silicon chip}}.
\newblock {\emph{\JournalTitle{Nature Photonics}}} \textbf{\bibinfo{volume}{4}}, \bibinfo{pages}{182--187} (\bibinfo{year}{2010}).

\bibitem{nanocavity1}
\bibinfo{author}{Nozaki, K.} \emph{et~al.}
\newblock \bibinfo{journal}{\bibinfo{title}{Ultralow-power all-optical ram based on nanocavities}}.
\newblock {\emph{\JournalTitle{Nature Photonics}}} \textbf{\bibinfo{volume}{6}}, \bibinfo{pages}{248--252} (\bibinfo{year}{2012}).

\bibitem{nanocavity2}
\bibinfo{author}{Kuramochi, E.} \emph{et~al.}
\newblock \bibinfo{journal}{\bibinfo{title}{Large-scale integration of wavelength-addressable all-optical memories on a photonic crystal chip}}.
\newblock {\emph{\JournalTitle{Nature Photonics}}} \textbf{\bibinfo{volume}{8}}, \bibinfo{pages}{474--481} (\bibinfo{year}{2014}).

\bibitem{nanocavity3}
\bibinfo{author}{Kuramochi, E.} \emph{et~al.}
\newblock \bibinfo{journal}{\bibinfo{title}{Ultralow bias power all-optical photonic crystal memory realized with systematically tuned l3 nanocavity}}.
\newblock {\emph{\JournalTitle{Applied Physics Letters}}} \textbf{\bibinfo{volume}{107}} (\bibinfo{year}{2015}).

\bibitem{iii_v_nanocavity1}
\bibinfo{author}{Alexoudi, T.} \& \bibinfo{author}{et~al.}
\newblock \bibinfo{journal}{\bibinfo{title}{Iii--v-on-si photonic crystal nanocavity laser technology for optical srams}}.
\newblock {\emph{\JournalTitle{IEEE Journal of Selected Topics in Quantum Electronics}}} \textbf{\bibinfo{volume}{22}}, \bibinfo{pages}{295--304} (\bibinfo{year}{2016}).

\bibitem{pcm_ram1}
\bibinfo{author}{R{\'\i}os, C.} \& \bibinfo{author}{et~al.}
\newblock \bibinfo{journal}{\bibinfo{title}{Integrated all-photonic non-volatile multi-level memory}}.
\newblock {\emph{\JournalTitle{Nature photonics}}} \textbf{\bibinfo{volume}{9}}, \bibinfo{pages}{725--732} (\bibinfo{year}{2015}).

\bibitem{pcm_ram2}
\bibinfo{author}{Pernice, W.~H.} \& \bibinfo{author}{Bhaskaran, H.}
\newblock \bibinfo{journal}{\bibinfo{title}{Photonic non-volatile memories using phase change materials}}.
\newblock {\emph{\JournalTitle{Applied Physics Letters}}} \textbf{\bibinfo{volume}{101}} (\bibinfo{year}{2012}).

\bibitem{pcm_ram3}
\bibinfo{author}{Miller, K.~J.}, \bibinfo{author}{Haglund~Jr, R.~F.} \& \bibinfo{author}{Weiss, S.~M.}
\newblock \bibinfo{journal}{\bibinfo{title}{Optical phase change materials in integrated silicon photonic devices}}.
\newblock {\emph{\JournalTitle{Optical Materials Express}}} \textbf{\bibinfo{volume}{8}}, \bibinfo{pages}{2415--2429} (\bibinfo{year}{2018}).

\bibitem{e_SRAM}
\bibinfo{author}{Karl, E.} \& \bibinfo{author}{et~al.}
\newblock \bibinfo{journal}{\bibinfo{title}{A 0.6 v, 1.5 ghz 84 mb sram in 14 nm finfet cmos technology with capacitive charge-sharing write assist circuitry}}.
\newblock {\emph{\JournalTitle{IEEE Journal of Solid-State Circuits}}} \textbf{\bibinfo{volume}{51}}, \bibinfo{pages}{222--229} (\bibinfo{year}{2015}).

\bibitem{photonic_imc_crossbar}
\bibinfo{author}{Ohno, S.}, \bibinfo{author}{Tang, R.}, \bibinfo{author}{Toprasertpong, K.}, \bibinfo{author}{Takagi, S.} \& \bibinfo{author}{Takenaka, M.}
\newblock \bibinfo{journal}{\bibinfo{title}{Si microring resonator crossbar array for on-chip inference and training of the optical neural network}}.
\newblock {\emph{\JournalTitle{Acs Photonics}}} \textbf{\bibinfo{volume}{9}}, \bibinfo{pages}{2614--2622} (\bibinfo{year}{2022}).

\bibitem{photonic_bnn}
\bibinfo{author}{Wang, R.} \emph{et~al.}
\newblock \bibinfo{journal}{\bibinfo{title}{Photonic binary convolutional neural network based on microring resonator array}}.
\newblock {\emph{\JournalTitle{IEEE Photonics Technology Letters}}} \textbf{\bibinfo{volume}{35}}, \bibinfo{pages}{664--667} (\bibinfo{year}{2023}).

\bibitem{photonic_imc_neuromorphic}
\bibinfo{author}{Tait, A.~N.} \emph{et~al.}
\newblock \bibinfo{journal}{\bibinfo{title}{Neuromorphic photonic networks using silicon photonic weight banks}}.
\newblock {\emph{\JournalTitle{Scientific reports}}} \textbf{\bibinfo{volume}{7}}, \bibinfo{pages}{7430} (\bibinfo{year}{2017}).

\bibitem{photonic_tensor_core}
\bibinfo{author}{Feldmann, J.} \emph{et~al.}
\newblock \bibinfo{journal}{\bibinfo{title}{Parallel convolutional processing using an integrated photonic tensor core}}.
\newblock {\emph{\JournalTitle{Nature}}} \textbf{\bibinfo{volume}{589}}, \bibinfo{pages}{52--58} (\bibinfo{year}{2021}).

\bibitem{photonic_sp}
\bibinfo{author}{Xie, Y.} \emph{et~al.}
\newblock \bibinfo{journal}{\bibinfo{title}{Towards large-scale programmable silicon photonic chip for signal processing}}.
\newblock {\emph{\JournalTitle{Nanophotonics}}} \textbf{\bibinfo{volume}{13}}, \bibinfo{pages}{2051--2073} (\bibinfo{year}{2024}).

\end{thebibliography}

\section*{Acknowledgments}
This work is supported by the Defense Advanced Research Projects Agency (DARPA) under Grant No. HR001123S0024. 


\end{document}